\documentclass[pra, 10pt, twocolumn, tightenlines, aps,superscriptaddress,amsmath,amssymb  nolongbibliography, nofootinbib]{revtex4-1}

\usepackage[utf8]{inputenc}
\usepackage[T1]{fontenc}
\usepackage[paper=a4paper,left=20mm,right=20mm,top=25mm,bottom=20mm]{geometry}
\usepackage{float}
\usepackage[english]{babel}
\usepackage[normalem]{ulem}
\usepackage{mathtools}
\usepackage{bbold}
\usepackage{graphicx}
\usepackage{multirow}
\usepackage{tabularx}
\usepackage[sgvnames, table]{xcolor}
\usepackage{soul}
\usepackage[colorlinks=true,allcolors=blue]{hyperref}

\DeclarePairedDelimiterX\braket[2]{\langle}{\rangle}{#1 \delimsize\vert #2}

\begin{document}
	
\title{Effective approaches to the dynamical properties \\ of two distinguishable Bose polarons}

\author{F. Theel}
\affiliation{Center for Optical Quantum Technologies, University of Hamburg, Department of Physics, Luruper Chaussee 149, D-22761, Hamburg, Germany}
\author{S. I. Mistakidis}
\affiliation{Department of Physics, Missouri University of Science and Technology, Rolla, MO 65409, USA}
\author{P. Schmelcher}
\affiliation{Center for Optical Quantum Technologies, University of Hamburg, Department of Physics, Luruper Chaussee 149, D-22761, Hamburg, Germany}
\affiliation{The Hamburg Centre for Ultrafast Imaging, University of Hamburg, Luruper Chaussee 149, D-22761, Hamburg, Germany}
	
\date{\today}

\begin{abstract} 
We explore effective approaches for describing the dynamics of induced interactions amongst two non-interacting distinguishable impurities - Bose polarons - when their couplings to the host species are switched on. First, we evaluate the time-dependent characteristics of each polaron through reduced single-particle models. Their validity is ultimately judged by comparing to observables obtained within a many-body variational approach. We argue that utilizing a time-dependent optimization process with the effective mass and frequency being fitting parameters leads to an accurate matching on the level of one-body observables with the prediction of the many-body method. 
Next, we construct two-body effective models, which besides the effective parameters of the single polaron, include a contact interaction potential modelling the induced interactions. 
It is shown that the instantaneous coupling strength obtained with a time-dependent optimization process allows for an accurate description of the impurities two-body correlation dynamics when compared to the many-body calculations. 
In this sense, it is possible to describe the dynamical crossover from correlated to anti-correlated impurity patterns related to the transition from attractive to repulsive induced interactions.  
Our results should facilitate the description of strongly particle-imbalanced mixtures via reduced models and hence their experimental understanding.
\end{abstract}
	
\maketitle

\section{Introduction}

Multicomponent ultracold quantum gases are versatile platforms to explore complex many-body states of matter \cite{bloch2008, cazalilla2011, langen2015}, since they exhibit an exquisite tunability in terms of system parameters such as species selective external potentials \cite{haas2007, henderson2009, barker2020}, interactions~\cite{papp2008, wu2011, hewitt2024} and atom number \cite{serwane2011, lester2018, sowinski2019}. 
In the case of strongly particle imbalanced two-component mixtures the minority species, called an impurity, is coupled to a majority gas and becomes dressed by its excitations forming a quasi-particle~\cite{landau1933, massignan2014, schmidt2018a} dubbed the Bose~\cite{spethmann2012, hu2016, jorgensen2016, skou2021, seetharam2021, grusdt2024} or Fermi~\cite{schirotzek2009, kohstall2012} polaron depending on the statistics of the host. Polaron properties are usually spectroscopically probed in the experiment \cite{mccabe2009, cetina2016}.

The large population imbalance naturally motivates reductions of the original many-body problem. These involve effective single-particle models~\cite{pastukhov2017, grusdt2017a, schmidt2018a, mistakidis2019a, jager2020} aiming to capture quasi-particle characteristics such as their effective mass and residue~\cite{massignan2014, penaardila2015, grusdt2017, jager2020}, or effective two-body approaches~\cite{naidon2018, huber2019, mistakidis2020a, mistakidis2020b} in order to describe induced impurity-impurity interactions mediated by the host.   
Specifically, the presence of attractive induced interaction has been identified when the impurities couple either both attractively or repulsively to their medium~\cite{recati2005, pavlov2018, dehkharghani2018, camacho-guardian2018a, reichert2019, reichert2019a, mistakidis2020a, mistakidis2020b, astrakharchik2023, paredes2024, grusdt2024}. This eventually leads for strong impurity-medium attractions to the generation of bound states, named bipolarons~\cite{casteels2013, camacho-guardian2018, will2021, jager2022}. 
However, another intriguing possibility is to consider a three-component setting where distinguishable impurities can couple with opposite signs to their bosonic host suggesting the existence of a mediated repulsive interaction potential~\cite{schecter2014, reichert2019, brauneis2021, theel2024}. In Ref. \cite{theel2024} the existence of mediated attractive and repulsive interactions was unveiled being  accompanied by a two-body correlated and anti-correlated behavior of the impurities, respectively.
A similar behavior has also been predicted for a fermionic host where the induced interaction can be approximated by a contact interaction potential \cite{de2014}.

Apparently, to reveal such processes three-component mixtures are required.
Relevant experimental realizations consist of isotopes including  ${}^{41}$K, ${}^{40}$K and ${}^{6}$Li~\cite{wu2011}, or ${}^{87}$Rb, ${}^{40}$K and ${}^{6}$Li~\cite{taglieber2008} as well as the three hyperfine states of spin-1 ${}^{87}$Rb condensates~\cite{lannig2020, bersano2018}. 
Notable investigations with three-component mixtures revolve from soliton creation and their interactions~\cite{lannig2020, bersano2018}, the formation of droplet-like configurations~\cite{ma2021, bighin2022, ma2023} to enriched lattice phases beyond the common superfluid to Mott-insulator transition~\cite{barman2015, liu2023}.
Other relevant studies investigated the impact of impurities on a binary mixture \cite{compagno2017, abdullaev2020, boudjemaa2020, keiler2021, becker2022, abbas2022}, thereby, focusing on either polaronic properties in a harmonic trap~\cite{boudjemaa2020, keiler2021}, or transport mechanisms within a double-well potential \cite{becker2022} as well as the interplay between a disordered external potential and the interspecies interactions in the emergent dynamical response \cite{abbas2022}.

The aim of the present work is to utilize the richness of three-component cold atom platforms in order to study the validity of effective approaches to capture dynamical properties of ensuing quasiparticles. 
The latter in our case are generated through embedding two non-interacting distinguishable impurities to a bosonic environment. 
Notice here also that the dynamical response of three-component systems is largely unexplored especially so for strongly imbalanced settings. Initially, we identify the ground state impurity-impurity induced correlation patterns with respect to the distinct impurity-medium interactions. 
As in Ref.~\cite{theel2024}, we showcase the existence of a crossover from attractive to repulsive induced interactions when the product of the different impurity-medium couplings changes sign.

Next, linear ramps of the impurity-medium interactions are performed across the distinct induced interaction regimes or remaining within the same. This process leads to a time-dependent effective mass and frequency of the polarons as well as induced correlations~\cite{mistakidis2020b,mistakidis2023}. 
To estimate these effective parameters we first construct different single-particle reduced models describing the composite system when solely one of the impurities interacts with the medium. 
To testify the applicability of these reduced approaches their predictions are directly compared with the ones of the correlated many-body simulations at the level of one-body observables.  
We find that the best agreement occurs for the model based on a time-dependent optimization using the effective mass and frequency as time-dependent fitting parameters for matching one-body observables within the many-body method. 
Other reductions, e.g. assuming adiabatic ramping of the interaction or that the host acts as an effective time-independent potential for the impurity, appear to be less accurate, e.g. due to neglecting the backaction of the impurities to the host. 

Subsequently, we develop two-body reduced models for quantifying the presence of induced correlations in the course of the evolution. These approaches inherit the information of the appropriate optimized one-body models and additionally contain an effective two-body interaction potential for the impurities.  
Their validity is confirmed by minimizing the deviation of the impurities two-body correlation function between the effective picture and the many-body simulations~\cite{mistakidis2020b}. 
Once again, we showcase that a time-dependent optimization procedure on the impurities effective coupling provides the most accurate description as compared to the many-body results. 
Interestingly, adequate (but less) agreement is observed by deploying an exponentially decaying interaction potential.
Finally, we quantify the presence of impurities-medium and impurity-impurity entanglement in the dynamics of the von Neumann entropy and the logarithmic negativity, respectively.

The present work is structured as follows. In Section~\ref{sec:setup} we introduce the many-body Hamiltonian of the three-component mixture and describe the applied linear ramp of the impurity-medium coupling. The main facets of the employed many-body variational method are explicated in Sec.~\ref{sec:methodology}. 
The ground state impurity-impurity induced correlation patterns are classified in Sec.~\ref{sec:ground_state}. 
The impurity dynamics triggered via linear ramps of the impurity-medium interactions is analyzed in detail in Sec.~\ref{sec:dynamics} along with the respective single- and two-particle reduced models used to quantify the time-dependent effective mass and frequency of the polaron.
We monitor the impurities-medium and impurities  entanglement in Sec.~\ref{sec:entanglement}.
We summarize our results and discuss possible future extensions in Sec.~ \ref{sec:conclusion}. 
In Appendix~\ref{app:ramp_time} we showcase the impact of the quench ramp rate. 
Appendix~\ref{app:eff_model_gs} elaborates on the reduced methods used to determine the ground state effective parameters, while in Appendix~\ref{app:opt_convergence} the convergence behavior of the optimization technique is explicated.
The applicability of the effective two-body model for various ramp protocols is demonstrated in Appendix~\ref{app:other_quench_protocols}. 
We estimate the dominant expansion coefficients of the impurities' correlation function in Appendix~\ref{app:decomposing_eff2b}.

%%%%%%%%%%%%%%%%%%%%%%%%%%%%%%%%%%%%%%%%%%%%%%%%%%%%%%%%%%%%%%%%%%%%%%%%%%%%%%%%%%%%%%%%%%%%%%%
\section{Three-component bosonic mixture}
\label{sec:setup}

We deploy an ultracold three-component atomic mixture in the presence of an external one-dimensional harmonic oscillator potential with $\omega_A=\omega_B=\omega_C \equiv \omega = 1$. 
Species $A$ consists of $N_A=10$ bosons and represents the host for the impurity species $B$ and $C$, where $N_B=N_C=1$. 
For simplicity, a mass-balanced system, i.e., $m_A=m_B=m_C \equiv m=1$, is assumed. 
Due to ultra-low temperatures it is adequate to use a contact s-wave interaction potential of strength $g_{\sigma,\sigma'}$, where $\sigma,\sigma'\in \{A,B,C\}$, for two atoms of the same or different species~\cite{chin2010}. It is known that the effective 1D interaction strengths,  $g_{\sigma\sigma'}$, depend on the respective three-dimensional scattering lengths $a_{\sigma\sigma'}^{3D}$ and the transverse confinement frequency $\omega_{\perp}$. 
The former can be tuned via Feshbach resonances~\cite{kohler2006, chin2010} and the latter leads to confinement induced resonances~\cite{olshanii1998}.  
The 1D nature of the emergent dynamics is ensured by utilizing an adequately large trap aspect ratio, e.g. $\omega_x/\omega_{\perp}=0.01$ customary in 1D experiments~\cite{lannig2020,romero2024experimental} in order to prevent involvement of transversal excitations.
Additionally, in view of the scope of the present work, namely to reveal the potential presence of induced impurity-impurity interactions,  we set $g_{BC}=0$. 
The many-body dimensionless Hamiltonian of this three-component setting is given by 
\begin{align}
\hat{H} =& \sum_{i=1}^{N_A} \hat{h}_{A}(x_i^A)  + g_{AA} \sum_{i<j} \delta(x_i^A - x_j^A) \nonumber \\
& + \hat{h}_B(x^B) + g_{AB} \sum_{i=1}^{N_A} \delta(x_i^A - x^B) \nonumber \\
& + \hat{h}_C(x^C) + g_{AC} \sum_{i=1}^{N_A} \delta(x_i^A - x^C).
\label{eq:MB_Hamiltonian}
\end{align}
Here, $\hat{h}_\sigma(x) = -(1/2)[ (d^2/dx^2) - x^2]$ denote the $\sigma$-component single-particle Hamiltonian terms. 
In the above Hamiltonian we employed harmonic oscillator units in which the energy scales as $\hbar\omega$ and the length and interactions are expressed in units of $\sqrt{\hbar/m\omega}$ and $\sqrt{\hbar^3\omega/m}$, respectively.
Experimentally, three-component mixtures have already been realized~\cite{taglieber2008, williams2009, wu2011, bersano2018} and thus our system should be within reach, at least, in terms of the considered population imbalance among the components. 
For instance, a promising experimental setup to utilize would consist of two Rubidium isotopes where, e.g., ${}^{85}$Rb represents the bath and two ${}^{87}$Rb atoms in different hyperfine states emulate the impurities~\cite{alvarez2013,egorov2013}. 
Similarly, employing a radiofrequency protocol as in Ref.~\cite{bersano2018} three different hyperfine states of $^{87}$Rb could be populated with the required imbalance in order to realize our system.

In the following, the three-component system will be prepared in different correlated ground state configurations with respect to the impurity-medium interaction strengths as outlined in Sec.~\ref{sec:ground_state}. Consecutively, a time-dependent linear ramp from an initial to a final value characterized by a ramping rate $\tau$ triggers the nonequilibrium dynamics, see Sec.~\ref{sec:dynamics}, aiming to quantify emergent impurity-impurity mediated interactions.
Specifically, the employed interaction ramp protocol reads 
\begin{align}
g_{A\sigma}(t) =
\begin{cases}
g_{A\sigma}^0 + \frac{g_{A\sigma}^\tau - g_{A\sigma}^0}{\tau}t, & t<\tau\\
g_{A\sigma}^\tau, & t \geq \tau, \\
\end{cases}
\label{eq:interaction_ramp}
\end{align}
where $g_{A\sigma}(t=0)\equiv g_{A\sigma}^{0}$ and $g_{A\sigma}(\tau\leq t)\equiv g_{A\sigma}^{\tau}$ with $\sigma=\{B,C\}$. Within the main text we shall mainly focus on ramp rates $\tau \omega = 1,5$. These ensure that the interaction ramp is slower compared to the  sudden quench and, thereby, leads to fewer high-energy excitations, but it is still far from being adiabatic [see also Appendix~\ref{app:ramp_time}].
Experimentally, ramping between two interaction strengths at different rates can be realized by adjusting the magnetic field strength near a corresponding Feshbach resonance. Hence, it is possible to tune the respective intra- or inter-component scattering lengths with a finite ramp rate, e.g. as in the experiments of Refs.~\cite{regal2003, desalvo2019}.

%%%%%%%%%%%%%%%%%%%%%%%%%%%%%%%%%%%%%%%%%%%%%%%%%%%%%%%%%%%%%%%%%%%%%%%%%%%%%%%%%%%%%%%%%%%%%%%
\section{Many-body computational method}
\label{sec:methodology}

To simulate the ground-state and the nonequilibrium quantum dynamics of the aforementioned three-component system we resort to the \textit{ab-initio} Multi-Layer Multi-configuration time-dependent Hartree method for atomic mixtures (ML-MCTDHX)~\cite{cao2013, kronke2013, cao2017}. 
In this context, the ansatz of the many-body wave function inherits a multi-layer structure with  time-dependent and variationally optimized basis functions such that all relevant intra- and inter-component correlations are taken into account.  In particular, the first truncation of the many-body wave function, $|\Psi(t)\rangle$, is performed on the so-called species layer. 
Namely, $|\Psi(t)\rangle$ is expanded in terms of ${D_\sigma}$ time-dependent species functions,  
$\{|\Psi_i^\sigma(t)\rangle\}_{i=1}^{D_\sigma}$ with $\sigma=A,B,C$, and time-dependent $C_{ijk}(t)$ coefficients, i.e.,
\begin{align}
|\Psi(t)\rangle = \sum_{ijk} C_{ijk}(t) |\Psi_i^A(t)\rangle \otimes |\Psi_j^B(t)\rangle \otimes |\Psi_k^C(t)\rangle.
\label{eq:Psi_MB}
\end{align}
Each $|\Psi_i^\sigma(t)\rangle$ describes the state of the entire $\sigma$ species, which is achieved by the second truncation on the so-called particle layer. This means that $|\Psi_i^\sigma(t)\rangle$ is expressed with respect to bosonic number states $|\vec{n}(t)\rangle$ with time-dependent expansion coefficients $C_{i,\vec{n}}^\sigma(t)$. 
The number states themselves are constructed from $d_\sigma$ time-dependent single-particle functions (SPFs), where the elements of $\vec{n}=(n_1, \dots,n_{d_\sigma})$ quantify the particle occupation in each SPF satisfying $\sum_i n_i=N_\sigma$. 
Eventually, each SPF is expanded into a time-independent discrete variable representation referring to a spatial grid~\cite{light1985}, that  contains in our case $\mathcal{M}_r=300$ grid points. 

Importantly, from the tensor coefficients $C_{ijk}(t)$ appearing in Eq.~(\ref{eq:Psi_MB}) the eigenvalues of the species reduced density matrices can be derived~\cite{cao2017}. 
They signify the contribution of each species function to the complete many-body wave function. Hence, they  provide information about the entanglement among the three different species~\cite{horodecki2009}, see also Sec.~\ref{sec:entanglement} for details about measuring bipartite entanglement between two subsystems.

Additionally, our ansatz is amenable to further reductions such that entanglement is ignored among all or specific components. 
As an extreme case example, if we deem to prohibit correlations among all species, then our ansatz in Eq.~(\ref{eq:Psi_MB}) contains $D_A=D_B=D_C=1$, leading to a single product state, i.e., $|\Psi(t)\rangle = |\Psi^A(t)\rangle \otimes |\Psi^B(t)\rangle \otimes |\Psi^C(t)\rangle$~\cite{pitaevskii2003, cao2017}.
This process is dubbed species mean-field \textit{sMF}, where each component acts as a mean-field type effective potential to the others but intracomponent correlations are present since $d_{\sigma} > 1$. 
This treatment is well-established for binary mixtures and has been used to uncover the effects stemming solely from interspecies correlations~\cite{mistakidis2019c, theel2020}. 

One feature that unveils the richness of the three-component mixture and renders it even more intriguing than the binary case is that it naturally allows for different types of entanglement which can be, for instance, explicated through distinct species mean-field ansatzes. 
For instance, let us consider a physical situation where entanglement between $A$ and $C$ subsystems is suppressed, but it still exists among $A$ and $B$. In this case, our ansatz should include just one species function for the $C$ impurity ($D_C=1$), while for species $A$ and $B$ it holds that $D_A=D_B>1$.
Below, this ansatz will be referred to as \textit{species mean-field for the $C$ component} (sMFC) since the latter cannot become entangled with the others. Here, from the perspective of the $AB$ subsystem, impurity $C$ acts as a mean-field type potential. 
Along the same lines, it is possible to define the respective sMFA and sMFB ansatzes. 

Finally, we remark that for our analysis additional calculations in terms of time-dependent effective one- and two-body models are performed. 
Indeed, for the single-particle model, described by $\hat{H}^{(1)}(x, t)$ [Sec.~\ref{sec:1b_model}], the dynamics is obtained by numerical integration of the time-dependent one-body Schr\"{o}dinger equation, i.e. $i\hbar\partial_t \Psi(x,t) = \hat{H}^{(1)}(x, t) \Psi(x,t)$. 
On the other hand, the time-dependent two-body problem is solved by expanding the two-body Hamiltonian $\hat{H}^{(2)}(x_1, x_2, t)$ [Sec.~\ref{sec:2b_model}] in terms of product states of one-body solutions.

\begin{figure}
\centering
\includegraphics[width=1.0\linewidth]{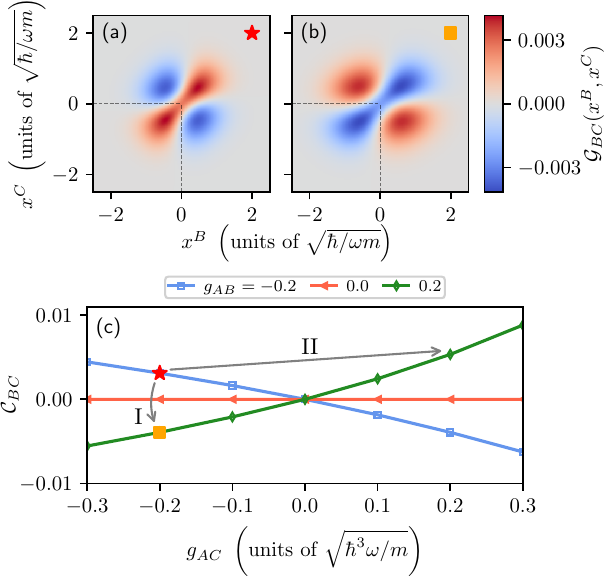}
\caption{(a), (b) Ground-state two-body correlation profiles, $\mathcal{G}_{BC}(x^B, x^C)$, of the impurities quantifying their induced correlations since $g_{BC}=0$. $\mathcal{G}_{BC}(x^B, x^C)$ shows the impurities are bunched (anti-bunched) at the same (different) position(s) as long as $g_{AB}g_{AC}>0$ ($g_{AB}g_{AC}<0$), see panel (a) for $(g_{AB}, g_{AC}) = (-0.2, -0.2)$, and panel (b) for $(g_{AB}, g_{AC}) = (-0.2, 0.2)$. 
(c) Integrating $\mathcal{G}_{BC}(x^B, x^C)$ over the region marked by the gray dashed lines in panels (a) and (b) provides information about the impurities two-body correlation content as quantified by $\mathcal{C}_{BC}$ for specific $g_{AB}$ (see legend) and varying $g_{AC}$.   
The gray arrows designated by $I$ and $II$ illustrate the two quench protocols performed later on in order to induce the dynamics.  
In all cases, the impurities are coupled to a bosonic medium containing $N_A=10$ atoms and  interacting with $g_{AA}=0.2$.
}
\label{fig:gs_corr_int}
\end{figure}

%%%%%%%%%%%%%%%%%%%%%%%%%%%%%%%%%%%%%%%%%%%%%%%%%%%%%%%%%%%%%%%%%%%%%%%%%%%%%%%%%%%%%%%%%%%%%%%
\section{Ground state correlation regimes}
\label{sec:ground_state}

Before delving into the dynamical behavior of the two impurities, we first provide a brief overview of the different ground-state correlation regimes which emerge upon varying the impurity-medium interaction strengths $g_{AB}$ and $g_{AC}$. 
More elaborated discussions and analysis on these phases can be found in Ref.~\cite{theel2024}.
In practice, two types of induced impurity-impurity correlations can be discerned. Namely, impurities a) bunching for $g_{AB}g_{AC}>0$ and b) anti-bunching occurring for $g_{AB}g_{AC}<0$. If $g_{AB}g_{AC}=0$, the $B$ and $C$ impurities remain uncorrelated. Recall that since $g_{BC}=0$, any signature of correlation among the impurities is induced from their interaction with the medium. 

A suitable observable to visualize this behavior is the spatially resolved two-body  correlation function~\cite{theel2021,theel2024} of the impurities 
\begin{align}
\mathcal{G}_{BC}(x^B, x^C) = \rho_{BC}^{(2)}(x^B, x^C) - \rho_B^{(1)}(x^B)\rho_C^{(1)}(x^C).
\label{eq:corr}
\end{align}
Here, the unconditional probability $\rho_B^{(1)}(x^B)\rho_C^{(1)}(x^C)$ constructed from the respective impurities one-body densities is subtracted from the conditional probability of finding the impurities at positions $x^B$ and $x^C$, i.e., the two-body reduced density matrix $\rho_{BC}^{(2)}(x_1^B, x_2^C)$~\cite{cao2017,mistakidis2020a}. This way, we are able to definitively identify the impurities two-body configurations.  
Specifically, the impurities located at $x^B$ and $x^C$ are two-body correlated (anti-correlated) as long as $\mathcal{G}_{BC}(x^B, x^C)>0$ ($\mathcal{G}_{BC}(x^B, x^C)<0$) and are uncorrelated for $\mathcal{G}_{BC}(x^B, x^C)=0$. 
Experimentally, the two-body correlation function can be observed in the following ways: i) detect single atoms after a time-of-flight expansion \cite{dall2013}, ii)  analyze single-shot images as described in Ref.~\cite{nguyen2019} or iii) through fluorescence measurements after freezing out the spatial degree's of freedom by ramping up a tight lattice confinement \cite{yao2024}.

In Figures~\ref{fig:gs_corr_int}(a) and (b) two characteristic correlation patterns of the impurities are presented for the interaction configurations $(g_{AB},g_{AC})=(-0.2, -0.2)$ and $(-0.2, 0.2)$,  respectively. 
It can be readily seen that in the case of $g_{AB}g_{AC}>0$ [Fig.~\ref{fig:gs_corr_int}(a)] the impurities are bunched namely they prefer to reside at the same location (see the diagonal of $\mathcal{G}_{BC}(x^B, x^C)$) and anti-bunched, i.e. they avoid each other, at different spatial regions (anti-diagonal of $\mathcal{G}_{BC}(x^B, x^C)$). 
In contrast, for $g_{AB}g_{AC}<0$ the impurity-impurity correlation pattern is reversed showing anti-bunching along the diagonal and bunching in the anti-diagonal, see Figure \ref{fig:gs_corr_int}(b).
Hence, the impurities tend to avoid each other at the same spatial regions but rather prefer to stay symmetrically placed with respect to the trap center. 
Such a crossover from induced correlation to anti-correlation has been independently analyzed in terms of an effective two-body model and found to be associated with either attractive or repulsive effective contact interactions in Ref.~\cite{theel2024}. 

To gain an overview of the impurities two-body correlation tendency upon different parametric variations, we subsequently measure the spatially integrated [see gray dashed lines in Fig.~\ref{fig:gs_corr_int}(a) and (b)] correlation function\footnote{While in the ground state the integration area is always either positive or negative, during the evolution it may contain both positive and negative values due to more complicated correlation patterns. However, we find that even in this case $\mathcal{C}_{BC}$ remains a useful tool as long as these effects are not dominant.}
\begin{align}
\mathcal{C}_{BC} = \int_{-\infty}^{0} \int_{-\infty}^{0} \mathrm{d}x^B \mathrm{d}x^C  \mathcal{G}_{BC}(x^B, x^C).
\label{eq:corr_int}
\end{align}
Thereby, a negative (positive) value of $\mathcal{C}_{BC}$ is associated with induced impurity-impurity anti-correlations (correlations), while $\mathcal{C}_{BC}=0$ indicates their absence. 
This observable is depicted in Fig.~\ref{fig:gs_corr_int}(c) for the ground state of the system as a function of $g_{AC}$ and for fixed values of $g_{AB}$. 
A crossover from impurities anti-bunching to bunching is captured once the sign of $g_{AB} g_{AC}$ is reversed as it was originally reported in Ref.~\cite{theel2024}.

\begin{figure*}
\centering
\includegraphics[width=1.0\linewidth]{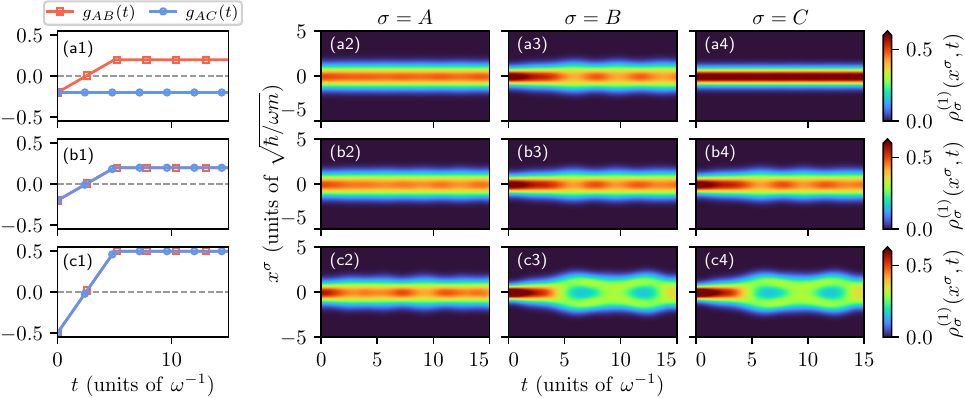}
\caption{Time-evolution of the $\sigma\in\{A,B,C\}$ component one-body densities, $\rho^{(1)}_\sigma(x,t)$, after linearly ramping up the impurity-medium interactions from $(g_{AB}^0, g_{AC}^0)$ to $(g_{AB}^\tau, g_{AC}^\tau)$ within $\tau \omega=5 $ as shown in panels (a1), (b1) and (c1). In particular, the interactions range (a1)-(a4) from $g_{AB}^0=-0.2$ to $g_{AB}^\tau=0.2$ while $g_{AC}^0=g_{AC}^\tau=-0.2$, (b1)-(b4) from  $g_{AB}^0=g_{AC}^0=-0.2$ to $g_{AB}^\tau=g_{AC}^\tau=0.2$ and (c1)-(c4) from $g_{AB}^0=g_{AC}^0=-0.5$ to $g_{AB}^\tau=g_{AC}^\tau=0.5$. The system is prepared in the ground state of two non-interacting $B$ and $C$ impurities immersed in a weakly coupled ($g_{AA}=0.2$) bosonic medium $A$ of $N_A=10$ atoms. The depicted one-body densities are given in units of $\sqrt{m\omega/\hbar}$.}
\label{fig:gpop_ramp}
\end{figure*}

%%%%%%%%%%%%%%%%%%%%%%%%%%%%%%%%%%%%%%%%%%%%%%%%%%%%%%%%%%%%%%%%%%%%%%%%%%%%%%%%%%%%%%%%%%%%%%%
\section{Dynamics of the distinguishable impurities after a  linear quench}
\label{sec:dynamics}

Having exemplified the interaction combinations ($g_{AB}$, $g_{AC}$) for which the impurities are predominantly correlated or anti-correlated in their ground-state, we aim to analyze the impurity dynamics after a time-dependent linear ramp of their intercomponent interactions. 
The latter are linearly modified at a specific rate $\tau$ from their initial attractive values ($g_{AB}^0$, $g_{AC}^0$) to final ones ($g_{AB}^\tau$, $g_{AC}^\tau$). 
Here, the overarching objective is to investigate the response of the impurity-impurity correlations when the related couplings are ramped from a correlated to an anti-correlated ground-state configuration as well as compare to scenarios at which the ramp is done between configurations with the same correlation behavior. 
For visualization purposes, these quench scenarios are illustrated in Fig.~\ref{fig:gs_corr_int}(c) with the gray arrows labeled with I and II.

\subsection{Density response}
\label{sec:1bd}

The time-evolution of the one-body density of component $\sigma$, $\rho^{(1)}_\sigma(x,t)$, after a linear interaction quench is presented in Fig.~\ref{fig:gpop_ramp}. 
Specifically, three paradigmatic linear interaction ramp protocols are utilized where either one or both impurity-medium interactions are ramped from attractive to repulsive values as depicted in Fig.~\ref{fig:gpop_ramp}(a1), (b1) and (c1) with a rate $\tau=5$. The latter value is adequately small to be close to the sudden quench and not large enough to approach the adiabatic limit and suppress excitation processes (cf. Appendix~\ref{app:ramp_time}).

An overall breathing motion of the cloud of each component is initiated by the intercomponent interaction ramp, as it can be seen from Fig.~\ref{fig:gpop_ramp} even though in some cases [e.g. Fig.~\ref{fig:gpop_ramp}(a2), (a4), (b2)] the breathing amplitude is relatively weak and thus the underlying collective motion becomes hardly visible \footnote{For a proper visualization of the breathing motion one can inspect, for instance, the variance of the one-body density (for brevity not shown here).}.
The breathing motion manifests by the collective, weak amplitude, expansion and contraction dynamics of the individual clouds. 
Specifically, for final impurity-medium interactions that do not exceed the intracomponent medium coupling as in Fig.~\ref{fig:gpop_ramp}(a1)-(a4) and (b1)-(b4) the components are miscible in the course of the evolution. Namely, they spatially overlap~\cite{timmermans1998, papp2008, pyzh2020} with the impurities remaining within the host, see in particular the spatial scales of the densities which are the same in all cases during the evolution.
This observation is important for the quasi-particle notion since, besides the formation of correlations, the finite overlap facilitates the impurity's dressing by the host excitations~\cite{rath2013, mistakidis2019a, mistakidis2019c}. 
Notice also that in the special case where $g_{AC}$ retains its original value (i.e. not quenched) the impurity $C$ is only slightly perturbed which is hardly visible see Fig.~\ref{fig:gpop_ramp}(a4). 
This is attributed to the indirect action of the impurity $B$ which through the finite change in $g_{AB}$ slightly modifies the distribution of the medium.
The dynamics of the latter for finite $g_{AC}$ eventually also perturbs impurity $C$. 

Turning to larger repulsive final impurity-medium interactions that overcome $g_{AA}$, following the quench,  we observe a somewhat altered impurity response as shown in Fig.~\ref{fig:gpop_ramp}(c2)-(c4). 
Here, dynamical miscibility is partially violated because the amplitude of the impurities ensuing breathing motion is arguably more pronounced due to stronger intercomponent repulsion. 
In this sense, at the time-instants of maximum expansion of the impurities breathing motion a partial immiscibility between the impurities and the host clouds takes place.
As such, temporal phase-separation events among the impurities and the medium take place while the impurities remain miscible throughout the evolution.  
Note in passing, that for even larger repulsive intercomponent interactions phase-separation becomes prominent and the impurities escape their host leading to a temporal orthogonality catastrophe of the quasi-particle picture~\cite{mistakidis2019b}. 
Still, in such repulsive interaction regimes the impurities, in the ground state of the system, can assemble outside of the bath in a Bell type distribution as it was recently shown in Ref.~\cite{anh-tai2024}.

To deepen our understanding on the above-described impurity response especially regarding the quasi-particles, i.e. Bose polarons, properties we subsequently construct effective descriptions to capture the dynamics of the two dressed impurities. To this end, we restrict ourselves to final intercomponent interactions being comparable with the intracomponent bath coupling $g_{AA}=0.2$, see in particular Fig.~\ref{fig:gpop_ramp}(a1)-(a4) and (b1)-(b4), such that the impurities reside within the medium.

%%%%%%%%%%%%%%%%%%%%%%%%%%%%%%%%%%%%%%%%%%%%%%%%%%%%%%%%%%%%%%%%%%%%%%%%%%%%%%%%%%%%%%%%%%%%%%%
\subsection{Effective one-body models}
\label{sec:1b_model}

As a first step, we develop the one-body building blocks which are proven to be essential for the effective two-body model to be discussed in Sec.~\ref{sec:2b_model}. 
The main idea is to compare the results obtained from the effective model for impurity $B$ ($C$) to the ones predicted by the appropriate sMF ansatz $|\Psi^{\mathrm{sMF}C}(t) \rangle$ ($|\Psi^{\mathrm{sMF}B}(t) \rangle$).
By doing so, the uncorrelated second impurity is seen by the other components only as a mean-field type potential. 
Recall that within the sMF$B$ (sMF$C$) ansatz only correlations between the medium and the impurity $B$ ($C$) are prohibited which is achieved by requiring $D_A,D_C>1~\textrm{and}~D_B=1$ ($D_A,D_B>1~\textrm{and}~D_C=1$) in the many-body wave function ansatz of Eq.~(\ref{eq:Psi_MB}).
We choose such sMF ansatzes for calculating the reference many-body wave function to which we will compare the effective model predictions. This choice is already justified by inspecting the corresponding densities evolution within the fully correlated system [Fig.~\ref{fig:gpop_ramp}(a3), (a4)] which are found to be qualitatively similar with  the sMF$B$ (sMF$C$) results (not shown for brevity). Thereby, we intentionally prohibit at this stage of the analysis the formation of induced correlations between the impurities since these effects are to be captured in a second step within the framework of an effective two-body model (Sec.~\ref{sec:2b_model}).

In the following, we introduce three different one-body models whose purpose is to effectively describe the dynamics of one impurity interacting with the remaining system. This is done exemplary for the scenario when both impurities are initially attractively coupled to the bath with $g_{AB}^0=g_{AC}^0=-0.2$ and then $g_{AB}$ is ramped to $g_{AB}^\tau=0.2$ within $\tau\omega=5$, while $g_{AC}(t)$ is held fixed in this case.

%------------------------------------------------------------------------------------------------------
\subsubsection{Effective mass and frequency with an adiabatic ramp}
\label{sec:eff1b_adiab}

\begin{figure*}
\centering
\includegraphics[width=1.0\linewidth]{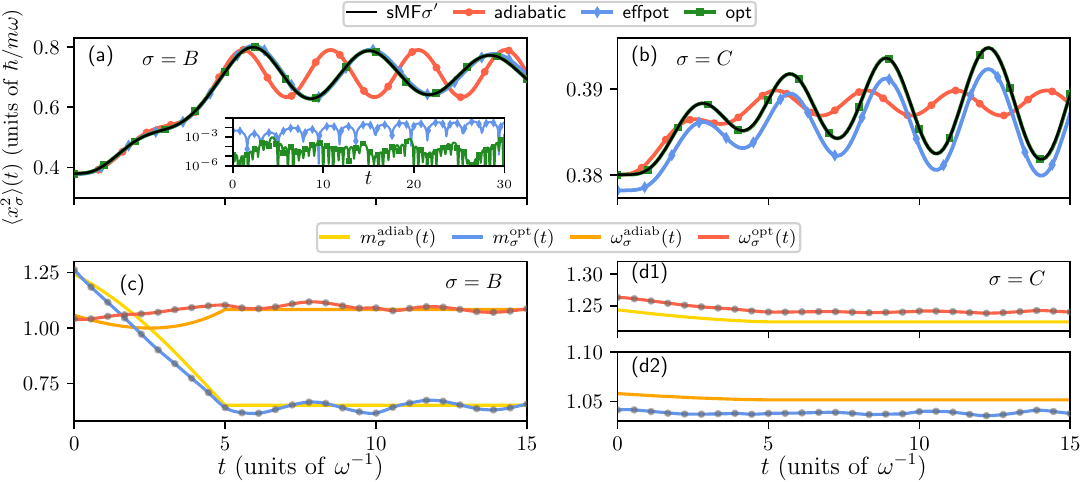}
\caption{Dynamics of the position variance for impurity (a) $B$ and (b) $C$ following a linear ramp of $g_{AB}(t)$ from $-0.2$ to $0.2$ within $\tau \omega=5$ while $g_{AC}=-0.2$ remains constant. The impurity dynamics within the sMF$\sigma'$, where $\sigma'=C,B$, is compared to three different effective one-body models (see legend) described in the main text. 
For $t \omega \geq 15$, the impurity's position variance obtained within the effective potential model (\textit{effpot}) starts to visibly deviate from the sMF$\sigma'$ prediction (not shown).
However, the variance following the optimization procedure (\textit{opt}) shows adequate agreement with the sMF$\sigma'$ outcome. For better visualization the inset of panel (a) depicts the relative differences between the variance of sMF$\sigma'$ and \textit{effpot} (green line) and \textit{opt} (blue line), namely $|\langle x_B^{(2)} \rangle_{\rm{sMF}\sigma'} -  \langle x_B^{(2)} \rangle_{\mathrm{effpot}}|/\langle x_B^{(2)} \rangle_{\rm{sMF}\sigma'}$ and $|\langle x_B^{(2)} \rangle_{\rm{sMF}\sigma'} -  \langle x_B^{(2)} \rangle_{\mathrm{opt}}|/\langle x_B^{(2)} \rangle_{\rm{sMF}\sigma'}$, respectively. 
Both impurities are initially coupled to a weakly interacting, $g_{AA}=0.2$, bosonic medium containing $N_A=10$ atoms. 
Time-dependent effective masses and frequencies of impurity (c) $B$ and (d1), (d2) $C$ used in the \textit{adiabatic} and \textit{optimized} models (see legend).}
\label{fig:1b_model_res}
\end{figure*}

Our first approach to effectively capture the impurity dynamics is to estimate for each time-dependent interaction configuration of the general form $(g_{AB}(t), g_{AC}(t))$, a corresponding effective mass ($m^{\rm{eff}}$) and frequency ($\omega^{\rm{eff}}$) from the respective ground states of the three-component setting, see Appendix~\ref{app:1b_model_gs} for details and also Ref.~\cite{theel2024}. 
In this way, it is possible to extract the interaction-dependent effective mass and frequency which will be referred to in the following as $m^{\rm{adiab}}(g_{AB}, g_{AC})$, $\omega^{\rm{adiab}}(g_{AB}, g_{AC})$. This means that these effective parameters practically describe the adiabatic polaron dynamics, namely the case where the interaction ramp is performed within long timescales ($\tau \gg 1$) compared to the characteristic ones of the system.  
As such, the time-dependent effective one-body model characterized by
$m_\sigma^{\rm{adiab}}(g_{AB}(t), g_{AC}(t)) \equiv m_\sigma^{\rm{adiab}}(t)$ and $\omega_\sigma^{\rm{adiab}}(g_{AB}(t), g_{AC}(t)) \equiv \omega_\sigma^{\rm{adiab}}(t)$, takes the form 
\begin{align}
\hat{H}_\sigma^{(1), \rm{adiab}}(t) &= -\frac{\partial_x^2}{2 m_\sigma^{\rm{adiab}}(t)} \nonumber \\
+& \frac{1}{2} m_\sigma^{\rm{adiab}}(t) \left(\omega_\sigma^{\rm{adiab}}(t)\right)^2 x^2.
\label{eq:eff1b_adiab}
\end{align} 
To testify the validity of $\hat{H}_\sigma^{(1), \rm{adiab}}(t)$, where $\sigma=B,C$, we invoke the impurities position variance $\langle x_\sigma^2\rangle(t)$ during the evolution. 
Specifically, the predictions of $\hat{H}_\sigma^{(1), \rm{adiab}}(t)$ and the corresponding sMF$\sigma'$ ansatz with $\sigma'=C,B$ are demonstrated in Fig.~\ref{fig:1b_model_res}(a), (b). 
For completeness, the behavior of $m_\sigma^{\rm{adiab}}(t)$ and $\omega_\sigma^{\rm{adiab}}(t)$ are also provided in Fig.~\ref{fig:1b_model_res}(c), (d1) and (d2). 

As it can be seen, a rough agreement between the two approaches (adiabatic and species mean-field) occurs on the level of the $C$ impurity variance [cf. Fig.~\ref{fig:1b_model_res}(b), note here the relatively small scales]. This is expected since only $g_{AB}$ is ramped resulting in a weak dynamical response of the $C$ component.
On the other hand, the adiabatic effective model fails to qualitatively estimate the dynamics of the $B$ impurity for $t\omega>5$. 
This deviation originates from the fact that the chosen ramping time is far from the adiabatic limit, see also the discussion in Appendix~\ref{app:ramp_time} about the impact of $\tau$ on the impurities dynamics. 
Interestingly, however, $\hat{H}_\sigma^{(1), \rm{adiab}}(t)$ is able to describe the overall trend of $\langle x_B^2\rangle(t)$ throughout the evolution.

%------------------------------------------------------------------------------------------------------
\subsubsection{Time-dependent effective one-body potential}
\label{sec:eff1b_effpot}

In the second effective model, we set $m_\sigma^{\mathrm{eff}}(t)=\omega_\sigma^{\mathrm{eff}}(t)=1$ and consider the mean-field type contribution of the bosonic medium to the bare one-body Hamiltonian $\hat{h}_\sigma$ appearing in Eq.~(\ref{eq:MB_Hamiltonian}). Here, the medium has the role of an external effective potential~\cite{mistakidis2019a, mistakidis2020a} and the resulting reduced model reads
\begin{align}
\hat{H}_\sigma^{(1), \rm{effpot}}(t) = \hat{h}_\sigma + g_{A\sigma}(t)N_A\rho_A^{(1),\mathrm{sMF}\sigma'}(x,t), 
\label{eq:eff1b_effpot}
\end{align}
where $\sigma=B,C$ and $\sigma'=C,B$. 
In particular, within this approach the one-body density of the medium is treated as a time-dependent mean-field potential which includes spatial deformations imprinted by the second impurity and it is weighted by the impurity-medium interaction strength $g_{A\sigma}(t)$. 
In the absence of interspecies correlations, i.e. for a full sMF ansatz, $\hat{H}_\sigma^{(1), \rm{effpot}}(t)$ models exactly the one-body dynamics of the impurities.

The temporal evolution of $\langle x_{\sigma}^2\rangle(t)$ obtained from $\hat{H}_\sigma^{(1), \rm{effpot}}(t)$ is presented in Fig.~\ref{fig:1b_model_res}(a), (b) for the impurity $B$ and $C$ respectively. 
We find that this model captures very well the dynamics of the impurities even though impurity-medium correlation effects are only indirectly taken into account by $\rho_A^{(1),\mathrm{sMF}\sigma'}(x,t)$ but are  otherwise neglected~\cite{keiler2019}.
This observation suggests that at these timescales and $g_{AB}$ interactions the impact of $AB$-entanglement on the spatial variance is suppressed.
However, for longer timescales, i.e. for $t\omega \sim 30$, we observe that the effective potential model can not adequately  capture the characteristics of the oscillatory behavior of $\langle x_{B}^2\rangle(t)$ obtained via the sMF$C$ method [see inset in Fig. \ref{fig:1b_model_res}(a)]. This discrepancy underlines the importance  of the impurity-medium correlation effects for the accurate description of the dynamics at least for longer timescales and for the present interaction quench.
In contrast, the agreement on $\langle x_{C}^2\rangle(t)$ is to a good degree anticipated since $g_{AC}$ retains its initial value. 
We have also confirmed that upon considering smaller ramping times, i.e. approaching the sudden quench scenario, the deviations among the predicted position variances take place at smaller timescales which is traced back to the larger amount of excitations induced by the quench.

%------------------------------------------------------------------------------------------------------
\subsubsection{Time-dependent optimization of the effective mass and frequency}\label{sec:eff1b_opt}

The structure of the third effective model corresponds to the one given by Eq.~(\ref{eq:eff1b_adiab}). However, instead of the effective mass and frequency measured from the (adiabatically followed) ground-state (or equivalently the adiabatic ramp), here they are acquired by following a time-dependent optimization method. 
Hence, $m_\sigma^{\rm{adiab}}(t)$ and $\omega_\sigma^{\rm{adiab}}(t)$ are replaced by $m_\sigma^{\rm{opt}}(t)$ and $\omega_\sigma^{\rm{opt}}(t)$ in Eq.~(\ref{eq:eff1b_adiab}). We refer to the respective optimized one-body Hamiltonian as $\hat{H}_{\sigma}^{(1), \rm{opt}}(t)$. 

The optimization process is initiated by solving the time-dependent one-body Schr\"{o}dinger equation for $\hat{H}_{\sigma}^{(1), \rm{opt}}(t)$ where the paths of $m_\sigma^{\rm{opt}}(t)$ and $\omega_\sigma^{\rm{opt}}(t)$ are dictated by a set of initial interpolation points. Meanwhile, a cost function $c_{\mathrm{opt}}^{1\mathrm{body}}$ (see below) is evaluated determining the quality of the effective wave function. Then, a gradient based optimization algorithm~\cite{byrd1995, zhu1997} determines a new set of interpolation points for the paths $m_\sigma^{\rm{opt}}(t)$ and $\omega_\sigma^{\rm{opt}}(t)$ leading to a new value of the cost function. This procedure is repeated until further varying the interpolation points does not improve the overlap of the effective and many-body wave functions [see also Appendix~\ref{app:opt_convergence}].
For our purposes, we choose the cost function  $c_{\mathrm{opt}}^{1\mathrm{body}} = \frac{1}{N_t}\sum_{i=1}^{N_t} \bigl[ |\langle x_\sigma^2\rangle_{\mathrm{sMF}\sigma'}(t_i) - \langle x^2\rangle_{\mathrm{eff}}(t_i)|/\langle x_\sigma^2\rangle_{\mathrm{sMF}\sigma'}(t_i) + \int \mathrm{d}x |\rho_{\sigma}^{(1),\mathrm{sMF}\sigma'}(x, t_i) - \rho_{\mathrm{eff}}^{(1)}(x, t_i)|^2 \bigl]$, where $t_{i+1} - t_i = 0.1$ denotes the time step and $N_t$ is the total number of timesteps. An optimization routine varies then the interpolation points such that the cost function becomes minimal.

By applying this procedure, we enforce the time-dependent wave function of the effective model to match the one of the many-body Hamiltonian (here extracted within the sMF$C$ method).
By design, this process leads to a quantitatively excellent agreement among the two approaches as exemplarily shown for the respective impurities position variances in Fig.~\ref{fig:1b_model_res}(a), (b). 
The extracted time-dependent behavior of $m_\sigma^{\rm{opt}}(t)$ and $\omega_\sigma^{\rm{opt}}(t)$ using $N_{\rm{interp}}=28$ is illustrated in Fig.~\ref{fig:1b_model_res}(c), (d1) and (d2). 
It becomes apparent that both $m_\sigma^{\rm{opt}}(t)$ and $\omega_\sigma^{\rm{opt}}(t)$ resemble to a certain degree $m_\sigma^{\rm{adiab}}(t)$ and $\omega_\sigma^{\rm{adiab}}(t)$ but they also exhibit additional time-dependent features caused by the quench and not captured by the adiabatic dynamics. 
In particular, the optimized parameters $m_\sigma^{\rm{opt}}(t)$ and $\omega_\sigma^{\rm{opt}}(t)$ depicted in Fig.~\ref{fig:1b_model_res}(c) fluctuate at times $t \geq \tau$ around the constant values of $m_\sigma^{\rm{adiab}}(t)$ and $\omega_\sigma^{\rm{adiab}}(t)$. These additional dynamical features of the optimized parameters unveil the non-adiabatic time-evolution of the system after the interaction ramp. They are intuitively expected since the interaction ramp imposes a breathing type motion on the different species (cf. Fig.~\ref{fig:gpop_ramp} where the complete many-body wave function has been employed) and thus time-dependent effective parameters are required for an accurate description.
Moreover, in the case of the $B$ impurity, and in both the adiabatic and optimized approaches, the effective mass decreases with time while the frequency slightly increases. 
This behavior is attributed to the presence of the harmonic trap and it is in accordance with Refs.~\cite{mistakidis2019b, mistakidis2019a}. Here, the post-quench repulsive impurity-medium coupling leads to a gradual delocalization of the impurity towards the trap edges in the course of the evolution. This can also be inferred from the apparent increase of the impurity's position variance. In this sense, the effective mass decreases~\cite{theel2024} since its dressing is reduced. 

On the other hand, a time-dependent quench from repulsive to attractive impurity-medium interactions leads to the inverse behavior, i.e., to the gradual increase of the effective mass during the evolution (not shown).  
Note in passing that in homogeneous setups an increase of effective mass should be expected as described in Refs.~\cite{nascimbene2009, penaardila2020}.
We also remark that this optimization method is able to effectively capture arising correlation effects emerging at longer timescales imprinted, for instance, as modulations in the time-dependent effective mass and frequency.

%%%%%%%%%%%%%%%%%%%%%%%%%%%%%%%%%%%%%%%%%%%%%%%%%%%%%%%%%%%%%%%%%%%%%%%%%%%%%%%%%%%%%%%%%%%%%%%
\subsection{Effective two-body models}
\label{sec:2b_model}

The success of the one-body effective description (especially with the optimization process) to estimate the effective mass and frequency of the generated polaron during the evolution motivates the consecutive investigation of the two non-interacting impurities dynamics. 
This process is, of course, more involved due to the occurrence of induced correlations and for this reason their effective interaction potential needs to be carefully chosen. 
As such, the effective two-body model has the form 
\begin{align}
\hat{H}^{(2),\rm{eff}}(t) = \hat{H}_{B}^{(1), \rm{opt}}(t) + \hat{H}_{C}^{(1), \rm{opt}}(t) + \hat{V}_{BC}^{\rm{int}}(t). 
\label{eq:2b_ham}
\end{align}
In this expression, the time-evolution of each polaron separately is described by $\hat{H}_{\sigma}^{(1), \rm{opt}}(t)$ incorporating $m_\sigma^{\rm{opt}}(t)$ and $\omega_\sigma^{\rm{opt}}(t)$, see also  Sec.~\ref{sec:eff1b_opt}, whilst induced impurity-impurity interactions are captured via the interaction potential $\hat{V}_{BC}^{\rm{int}}(t)$. 
Below, we elaborate on the applicability of three different forms of this time-dependent two-body effective potential by comparing with many-body results obtained within the full many-body ML-MCTDHX method where all correlations (including induced two-body ones) are considered.
To validate the choice of $\hat{V}_{BC}^{\rm{int}}(t)$ we employ, as case examples, two quench protocols. Namely, the first facilitates the crossover from initially correlated impurities to anti-correlated ones [cf. arrow $I$ in Fig.~\ref{fig:gs_corr_int}(c)], while in the second case we ramp the impurity-medium interactions between two correlated states [cf. arrow $II$ in Fig.~\ref{fig:gs_corr_int}(c)].

To justify the necessity of the inclusion of the two-body interaction potential we compare the many-body ML-MCTDHX results to the solutions of $\hat{H}^{(2),\rm{eff}}(t)$ with $\hat{V}_{BC}^{\rm{int}}(t)=0$. 
For this purpose, we use two representative two-body observables. 
The first is the impurities relative distance~\cite{mistakidis2019, mistakidis2020a} defined by 
\begin{align}
\langle r_{BC} \rangle(t) &= \frac{1}{N_BN_C} \nonumber \\
\times \int &{\mathrm{d}} x^B {\mathrm{d}} x^C \left|x^B - x^C \right| \rho_{BC}^{(2)}(x^B, x^C, t),
\label{eq:rel_dist}
\end{align}
where the corresponding dynamics for both of the above-described quench protocols and approaches is presented in Fig.~\ref{fig:2b_model_res}(a1), (b1). 
It can be readily seen that the uncoupled model adequately captures the dynamics of $\langle r_{BC} \rangle(t)$ implying that the impact of two-body effects is comparatively small at this level. 
However, this picture changes drastically when one inspects the integrated impurity-impurity correlation function, $\mathcal{C}^{\rm{int}}_{BC}(t)$, see Fig.~\ref{fig:2b_model_res}(a2), (b2). Indeed, since interparticle correlations are vanishing within the uncoupled model it holds that $\mathcal{C}^{\rm{int}}_{BC}(t)=0$, which is in contrast to the predictions of the fully correlated approach. In this latter case, the evolution of $\mathcal{C}^{\rm{int}}_{BC}(t)$ reflects the characteristics of the pre- and post-quench ground states. 
In fact, $\mathcal{C}^{\rm{int}}_{BC}(t)$ transits from a correlated to an anti-correlated behavior for the first quench [Fig.~\ref{fig:2b_model_res}(a2)], while it remains positive in the second quench scenario [Fig.~\ref{fig:2b_model_res}(b2)] evincing a robust correlated trend.  
In this sense, these observations set the stage for searching a suitable two-body interaction potential.

\subsubsection{Two-body potential with effective interactions from the adiabatic ramp}
\label{sec:eff2b_adiab}

Here, $\hat{V}_{BC}^{\rm{int}}(t)$ contains an effective impurity-impurity coupling that is estimated (as in Sec.~\ref{sec:eff1b_adiab}) from each ground state configuration  encountered along the corresponding adiabatic path of the linear ramp protocol. 
To determine these effective couplings, $g_{BC}^{\mathrm{adiab}}(g_{AB}, g_{AC})$, we demand the matching of the two-body correlation functions among the static effective model and the ground state (or equivalently the adiabatic solution) of the many-body method, see details in Appendix~\ref{app:2b_model_gs}. 
In this sense, we refer to the effective two-body model based on the adiabatic approximation as $\hat{H}^{(2),\rm{adiab}}(t)$ and set $\hat{V}_{BC}^{\rm{int}}(t)=g_{BC}^{\mathrm{adiab}}(t)\delta(x^B - x^C)$.

\begin{figure*}
\centering
\includegraphics[width=0.95\linewidth]{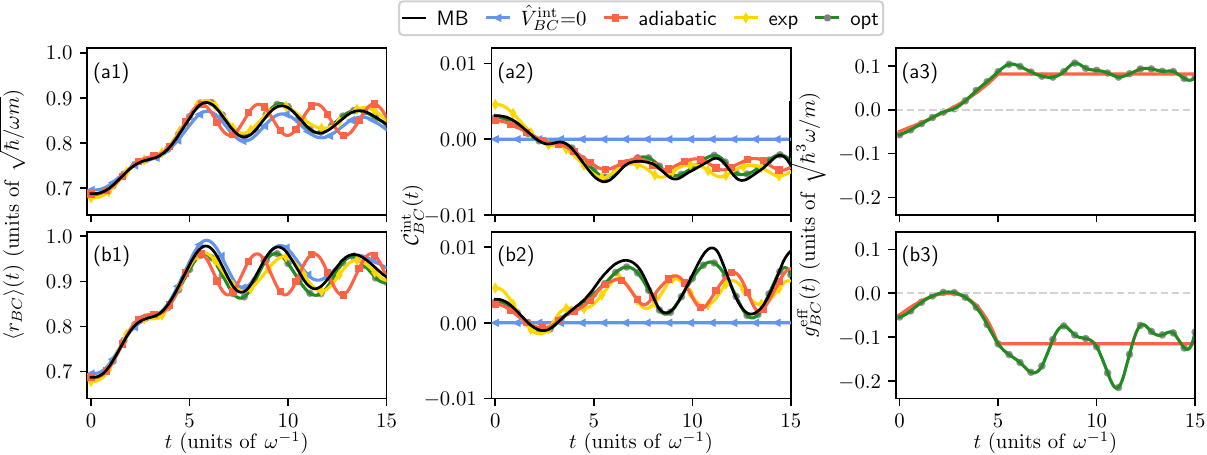}
\caption{Comparison of three effective two-body models (see main text) to the many-body dynamics (labeled as \textit{MB}) of two distinguishable and non-interacting impurities $B$ and $C$ coupled to a weakly interacting bosonic medium with $N_A=10$ atoms and $g_{AA}=0.2$. 
The impurities couple initially attractively to the bath with $(g_{AB}^0, g_{AC}^0)=(-0.2, -0.2)$ and are then linearly ramped within $\tau\omega=5$ either towards $(g_{AB}^\tau, g_{AC}^\tau)=(0.2, -0.2)$ (top row) or to $(g_{AB}^\tau, g_{AC}^\tau=(0.2, 0.2)$ (bottom row) [cf. arrows in Figure \ref{fig:gs_corr_int}(c)]. The prediction of the different methods are compared with respect to the impurities (a1), (b1) relative distance and (a2), (b2) integrated correlation function. 
The latter features a transition from a correlated to an anti-correlated behavior following the first quench and it remains correlated for the second quench. 
The time-dependent effective interaction strength used in the \textit{adiabatic} and the \textit{optimized} models are shown in panels (a3), (b3). The gray circles denote the interpolation points of the optimization method and the gray dashed lines mark the non-interacting case, i.e., when $g_{BC}^{\mathrm{eff}}=0$.}
\label{fig:2b_model_res}
\end{figure*}

The quench-induced dynamics of either $\langle r_{BC} \rangle(t)$ [Fig.~\ref{fig:2b_model_res}(a1), (b1)] or  $\mathcal{C}^{\rm{int}}_{BC}(t)$ [Fig.~\ref{fig:2b_model_res}(a2), (b2)] shows an adequate agreement between the effective approach and the many-body method for $t \lesssim \tau$. However, deviations exist at later evolution times especially so in the case that both impurity-medium interactions are ramped-up. These aberrations partly stem from the fact that the applied interaction ramp cannot be considered to be adiabatic. 
Therefore, also the rather complex dynamics of the induced correlations after the quench is unlikely to be well captured by $g_{BC}^{\mathrm{adiab}}(t)$, which for these times is constant [Fig.~\ref{fig:2b_model_res}(a3), (b3)]. 
It is also interesting to note here that the effective interaction exhibits a complementary behavior to the correlation function. For instance, as expected from the ground state, attractive effective interactions go in sync with correlated two-body behavior and vice versa.

\subsubsection{Exponential interaction potential}
\label{sec:eff2b_exp}

As a second attempt for modelling the impurities effective two-body interaction potential we consider an exponential Yukawa-type potential. Such an interaction potential has been used to describe the induced interaction in the ground state of polarons with a homogeneous background in Refs.~\cite{recati2005, reichert2019, reichert2019a, will2021, grusdt2024}, but has also been recently applied to the case of a harmonically trapped medium~\cite{theel2024}. 
It reads 
\begin{align}
\hat{V}_{BC}^{\rm{int}}(r_{BC}, t) = -\frac{g_{AB}(t)g_{AC}(t)m_A}{\sqrt{\gamma}}e^{-2r_{BC}/\xi_A},
\label{eq:ind_int_yukawa}
\end{align}
where $r_{BC}=|x^B-x^C|$ denotes the impurities' distance and $\gamma=m_Ag_{AA}/(N_A\rho_A^{(1)}(0))$. 
Here, the time-dependence is inherited by the linearly ramped coupling parameters $g_{AB}(t)$ and $g_{AC}(t)$.
Furthermore, it is required that $r_{BC} \lesssim \xi_A$ meaning that the impurities' distance should be comparable to the healing length of the medium $\xi_A \approx 1/\sqrt{2m_Ag_{AA}N_A\rho_A^{(1)}(0)}\approx 0.7$. 

Substituting $\hat{V}_{BC}^{\rm{int}}(r_{BC}, t)$ in Eq.~(\ref{eq:2b_ham}) we construct the effective two-body Hamiltonian dubbed  $\hat{H}^{(2),\rm{exp}}(t)$. 
The latter is numerically solved in order to extract $\langle r_{BC} \rangle(t)$ and $\mathcal{C}^{\rm{int}}_{BC}(t)$ which are subsequently compared to the predictions of the ML-MCTDHX method, see Fig.~\ref{fig:2b_model_res}(a1), (b1).
It can be deduced that $\hat{H}^{(2),\rm{exp}}(t)$ is indeed able to qualitatively describe the evolution of the impurities' relative distance for both quenches much better than $\hat{H}^{(2),\rm{adiab}}(t)$. 
Regarding $\mathcal{C}^{\rm{int}}_{BC}(t)$, the $\hat{H}^{(2),\rm{exp}}(t)$ approach shows a similar behavior as  $\hat{H}^{(2),\rm{adiab}}(t)$.
This means a qualitative good agreement with the many-body evolution  for $t\lesssim\tau$ and a growing deviation at longer times. Still, it appears that the nonlocal character of the applied interaction potential plays a crucial role and captures the oscillation frequency of the relative distance during the evolution much  better when compared to the adiabatic approach. Deviations observed for longer times are partially traced back to the fact that after the quench this potential is static and hence cannot account for dynamical deformations that exist.

\subsubsection{Time-dependent optimization of the contact interaction potential}
\label{sec:eff2b_opt}

The third approach to estimate the effective strength of the impurities induced interactions is to follow an optimization procedure similar to the one used in Sec.~\ref{sec:eff1b_opt} for identifying the polaron effective mass and frequency. This will allow to determine the optimal time-dependent induced coupling, $g_{BC}^{\mathrm{opt}}(t)$, by minimizing the cost function of the two-body impurity correlations. 
Here, the effective Hamiltonian $\hat{H}^{(2),\rm{opt}}(t)$ [Eq.~(\ref{eq:2b_ham})] encapsulates the interaction potential $\hat{V}_{BC}^{\rm{int}}(t)=g_{BC}^{\mathrm{opt}}(t)\delta(x^B - x^C)$, where $g_{BC}^{\mathrm{opt}}(t)$ is represented by a finite set of interpolation points $N_{\rm{interp}}$ \footnote{$N_{\rm{interp}}$ determines the accuracy of $g_{BC}^{\mathrm{opt}}(t)$ as briefly discussed in Appendix~\ref{app:opt_convergence}.}. 
Specifically, considering an initial set of equidistant interpolation points we evaluate the cost function  $c_{\mathrm{opt}}^{2\mathrm{body}}= \frac{1}{N_t}\sum_{i=1}^{N_t} \int \int \rm{d}x^B\rm{d}x^C | \mathcal{G}_{BC}(x^B, x^C, t_i) - \mathcal{G}_{BC}^{\rm{eff,opt}}(x^B, x^C, t_i)|$ with $\mathcal{G}_{BC}(x^B, x^C, t_i)$ obtained from the full many-body approach. 
Note that the timesteps $t_{i+1} - t_i = 0.1$ and $N_t$ is the total number of timesteps. 
Based on the outcome of the value of the cost function a gradient-based optimization routine varies the amplitudes of the time-wise fixed number of interpolation points such that $c_{\mathrm{opt}}^{2\mathrm{body}}$ is minimized. Namely, upon further varying the interpolation points does not lead to a smaller cost value.

The optimized $g_{BC}^{\mathrm{opt}}(t)$ is shown in Fig.~\ref{fig:2b_model_res}(a3), (b3) together with $N_{\rm{interp}}=28$ (gray dots). 
It is evident that the optimized path of $g_{BC}^{\mathrm{opt}}(t)$ agrees well with $g_{BC}^{\mathrm{adiab}}(t)$ for $t\lesssim\tau$. This reflects the equally good description of $\langle r_{BC} \rangle(t)$ and $\mathcal{C}^{\rm{int}}_{BC}(t)$ within this time interval [see Fig.~\ref{fig:2b_model_res}(a1), (b1) and (a2), (b2), respectively]. 
Turning to $t>\tau$, where $\mathcal{C}_{BC}^{\mathrm{int}}(t)$ features a more complex behavior only the optimized effective description is able to correctly describe the impurities correlation dynamics since the other effective models (as argued above) neglect certain correlation channels. 
At this point, it would be instructive to remark that we have also checked the performance of the optimization with respect to a time-dependent exponential interaction potential of the form of Eq.~(\ref{eq:ind_int_yukawa}). In this case, both the factor and the exponent of the Yukawa-type interaction potential have been optimized. Thereby, we found that the respective two-body model yields a similar agreement with the complete many-body ansatz predictions as compared to the effective model characterized by an optimized contact interaction potential.
This observation indicates that accounting for the exponential tails of the interaction potential is not decisive for a quantitatively accurate description in the harmonically trapped case. Instead, one has to deploy interaction potentials which depend on both impurity coordinates and not solely on their relative difference \cite{dehkharghani2018, chen2018}.

Summarizing, it was found that the optimization scheme is able to adequately describe the induced correlation dynamics of the impurities throughout the evolution. 
An intriguing feature of the time-dependent effective coupling strength is that it exhibits a crossover from attractive to repulsive induced interactions (already known on the ground state level~\cite{theel2024}). This can explain the observed transition of the impurities' two-body correlation from a correlated to an anti-correlated behavior.
Another imprint of this interesting induced correlation aspect is the dynamical evolution of the bipartite entanglement, see Sec.~\ref{sec:entanglement} about similarities of the von Neumann entropy involving the medium and the impurities entanglement.
Furthermore, it is interesting to  note that a shorter ramp rate may trigger a repeated crossing from induced attraction to repulsion and vice versa. Such a case is investigated in more detail in Appendix~\ref{app:decomposing_eff2b} elaborating also on the expansion coefficients of the impurities' correlation function.

\begin{figure}
\centering
\includegraphics[width=1.0\linewidth]{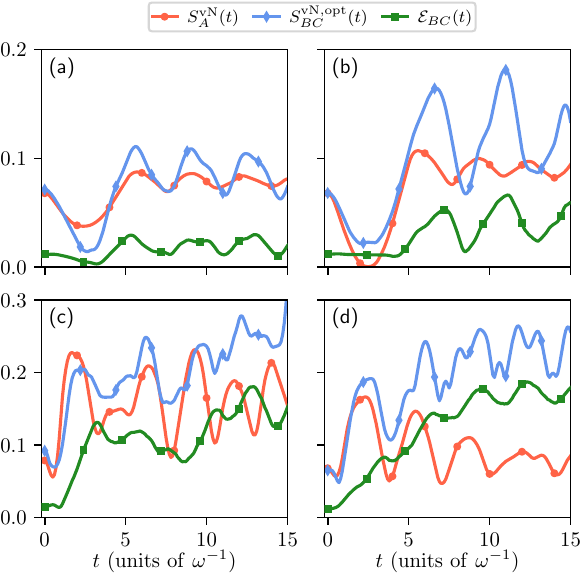}
\caption{Dynamical response of the von Neumann entropy quantifying entanglement among the medium and the two impurities subsystems, $S_A^{\mathrm{vN}}(t)$, the impurities' logarithmic negativity, $\mathcal{E}_{BC}(t)$, and the von Neumann entropy obtained from the effective two-body wave function, $S_{BC}^{\mathrm{vN, opt}}(t)$ (see also Sec.~\ref{sec:entanglement} for further details on the extraction of these observables). Independently of the measure, the impurity-impurity entanglement is finite justifying the presence of induced interactions. To induce the dynamics the impurity-medium interaction strengths are linearly ramped (a) from $(g_{AB}^0, g_{AC}^0)=(-0.2, -0.2)$ to $(g_{AB}^\tau, g_{AC}^\tau)=(0.2, -0.2)$ with rate $\tau=5$, (b) from $(g_{AB}^0, g_{AC}^0)=(-0.2, -0.2)$ to $(g_{AB}^\tau, g_{AC}^\tau)=(0.2, 0.2)$ with $\tau=5$, (c) from $(g_{AB}^0, g_{AC}^0)=(-0.2, 0.2)$ to $(g_{AB}^\tau, g_{AC}^\tau)=(0.2, -0.2)$ using $\tau=1$ and (d) from $(g_{AB}^0, g_{AC}^0)=(-0.2, -0.2)$ to $(g_{AB}^\tau, g_{AC}^\tau)=(-0.2, 0.2)$ with $\tau=1$.}
\label{fig:entanglement}
\end{figure}

\section{Dynamics of bipartite entanglement measures}
\label{sec:entanglement}

The existence of bipartite entanglement among the impurities subsystem with the medium is essential for the build-up of impurities induced correlations mediated by their host.
A frequently used measure for estimating the degree of bipartite entanglement of pure states is the von-Neumann entropy~\cite{horodecki2009}. In order to calculate the von Neumann entropy between the medium and the two distinguishable impurities, we express the many-body wave function given by Eq.~(\ref{eq:Psi_MB}) as a truncated Schmidt decomposition~\cite{schmidt1907, ekert1995} of rank $D_s=\min\{D_A, D_B\cdot D_C\}$. Namely 
\begin{align}
|\Psi(t)\rangle = \sum_{i=1}^{D_s} \sqrt{\lambda_i(t)} |\tilde{\Psi}_i^A\rangle \otimes |\tilde{\Psi}_i^{BC}\rangle, 
\end{align}
where $\lambda_i$ are the so-called Schmidt coefficients. The latter are the crucial ingredients of the von-Neumann entropy which reads 
\begin{align}
S^{\mathrm{vN}}_{A}(t) = - \sum_{i=1}^{D_s} \lambda_i(t) \ln \lambda_i(t).
\label{eq:von_neumann_entropy}
\end{align}
Apparently, $S^{\mathrm{vN}}_{A}(t)$ (or equivalently entanglement) becomes maximal within our Hilbert space truncation if all Schmidt coefficients are equally populated, i.e., $\lambda_i=1/D_s$. 
Otherwise, $S^{\mathrm{vN}}_{A}(t)$ vanishes if only one Schmidt-coefficient is non-zero implying that the two-subsystems are described by a product state.
It turns out that due to the hierarchical ordering of the time-dependent Schmidt-coefficients for the considered dynamical cases in our setup (not shown for brevity) the maximally allowed entropy (due to truncation) is never reached. In fact, its upper bound lies well above the observed values of the von Neumann entropy shown in Fig.~\ref{fig:entanglement}.

The time-evolution of $S^{\mathrm{vN}}_{A}(t)$ is presented in Fig.~\ref{fig:entanglement} where the upper panels show the entanglement following the time-dependent quench protocols discussed in Sec.~\ref{sec:eff2b_opt} and the lower panels refer to the protocols illustrated in Fig.~\ref{fig:other_quenches_entanglement}(a1), (b1) discussed in Appendix~\ref{app:other_quench_protocols}.
It can be seen that the impurities-medium entanglement [as captured through $S^{\mathrm{vN}}_{A}(t)$] is finite in all cases even at $t=0$, thus justifying the presence of impurity-impurity induced correlations already from the ground state of the system but also evincing their systematic build-up during the dynamics.  
Notice also here, for completeness, that the response of $S^{\mathrm{vN}}_{A}(t)$  partially follows the time-evolution of the medium's variance $\langle x_A^2 \rangle(t)$ and the mean relative distance between the impurities and the medium $(\langle r_{AB} \rangle(t) + \langle r_{AC} \rangle(t))/2$ (not shown).

On the other hand, instead of exploiting the impurities integrated correlation function (in order to infer the nature of their induced correlations) another possibility to quantify the impurities' entanglement is represented by the so-called logarithmic negativity~\cite{vidal2002} denoting an entanglement monotone \cite{plenio2005}.
It measures the bipartite entanglement of a mixed state and has been already successfully applied e.g. in Refs.~\cite{horhammer2008,duarte2009,zell2009,shiokawa2009,charalambous2019,becker2022,theel2024}.
To compute this observable, we first trace out the medium's (subsystem $A$) degrees of freedom from $| \Psi (t)\rangle \langle \Psi(t)|$. This process leads to a two-component density matrix describing the impurities' subsystem
\begin{subequations}
\begin{align}
\rho^{(2),\mathrm{spec}}_{BC}(t) =& \mathrm{Tr}_A \left( | \Psi (t)\rangle \langle \Psi(t)| \right) \\
= \sum_{jkmn}\sum_i & C_{ijk}(t) C_{imn}^*(t) \nonumber \\
\times |\Psi^B_j(t) \rangle&  \langle \Psi^B_m(t)| \otimes |\Psi^C_k(t) \rangle \langle \Psi^C_n(t)|,
\end{align}
\end{subequations}
where $|\Psi^B_j \rangle$ and $|\Psi^C_k \rangle$ denote the species functions of impurity $B$ and $C$, respectively, while $C_{ijk}(t)$ are the time-dependent expansion coefficients. 
If the impurity $B$ is not entangled with $C$ then the two-component mixture is separable and the eigenvalues of $\rho^{(2),\mathrm{spec}}_{BC}(t)$ are always positive. This is known as the positive partial transpose (PPT) criterion~\cite{peres1996, horodecki2009}.
As such, one pathway to reveal the presence of entanglement is to search for negative eigenvalues \footnote{Note that the reverse case, i.e., the presence of entanglement leads to negative eigenvalues of $\rho^{(2),\mathrm{spec}}_{BC}(t)$, is only true for small dimensions, namely ($2\times2$) or ($2\times3$), of $\rho^{(2),\mathrm{spec}}_{BC}(t)$ which does not apply here~\cite{peres1996}.} in the spectrum of the partially transposed two-component density matrix, $\left(\rho^{(2),\mathrm{spec}}_{BC}(t)|_{jkmn} \right)^{T_B}=\rho^{(2),\mathrm{spec}}_{BC}(t)|_{mkjn}$~\cite{zyczkowski1998,braun2002}.
Accordingly, the presence of bipartite entanglement is testified by the logarithmic negativity 
\begin{align}
\mathcal{E}_{BC} (t) = \log_2(1+2\mathcal{N}(t)),
\end{align}
where $\mathcal{N}$ refers to the sum of the negative eigenvalues of $\rho^{(2),\mathrm{spec}}_{BC}(t)$ multiplied by $(-1)$.

An alternative way to estimate impurities bipartite entanglement is via the von Neumann entropy that is calculated from the two-body wave function obtained by solving the effective two-body model within the optimization procedure as described in Sec.~\ref{sec:eff2b_opt}. The resulting time-dependent wave function is a pure state from which the von Neumann entropy denoted by $S^{\rm{vN,opt}}_{BC}(t)$ can be extracted via Eq.~(\ref{eq:von_neumann_entropy}).
The two measures describing the impurities  induced entanglement, i.e., $\mathcal{E}_{BC}(t)$ and $S^{\rm{vN,opt}}_{BC}(t)$ are shown in Fig.~\ref{fig:entanglement}. 
We observe that both measures exhibit a qualitative similar oscillatory dynamics and are, therefore, both capable of providing adequate information about the induced bipartite entanglement between the impurities. 
Another interesting observation can be made by comparing the dynamics of $S^{\mathrm{vN}}_{A}(t)$, i.e. the entanglement between the medium and the impurities, with the behavior of the impurity-impurity entanglement captured by $S^{\rm{vN,opt}}_{BC}(t)$ and $\mathcal{E}_{BC}(t)$ for the interaction ramps presented in  Fig.~\ref{fig:entanglement}(a)-(c). It is evident that an increase (decrease) of the induced entanglement among the impurities is accompanied by a similar increasing (decreasing) trend of the impurity-medium correlations. 
This hints to an indirect relation between the respective types of entanglement.
On the other hand, in Fig.~\ref{fig:entanglement}(d) we find opposite trends between the impurities-medium and the impurity-impurity entanglements. This can be attributed, in part, to the fact that in this case the impurity-medium interaction remains intact.

We also remark that a similar qualitative behavior between $S^{\mathrm{vN}}_{A}(t)$ and the impurity-medium relative distance $(\langle r_{AB} \rangle(t) + \langle r_{AC} \rangle(t))/2$ as well as among $S^{\rm{vN,opt}}_{BC}(t)$ and the impurities' relative distance $\langle r_{BC} \rangle(t)$ has been observed for times $t>\tau$ (not shown for brevity).
However, in order to develop a better understanding on the link between the dynamical response of $S^{\mathrm{vN}}_{A}(t)$ and $S^{\rm{vN,opt}}_{BC}(t)$ with the aforementioned relative distances more systematic studies (both numerically covering the available parametric space and analytically whenever possible) on the relevant measures are certainly required. 
Along the same lines, another interesting future research direction would be the systematic study of the interplay between the impurity-medium entanglement and the induced impurity-impurity entanglement.

\section{Conclusions and perspectives}
\label{sec:conclusion}

We have studied the validity of various one- and two-body effective models for describing the dynamical response of two non-interacting distinguishable impurities immersed in a bosonic medium. 
This is achieved by a direct comparison to the predictions of a full many-body numerical approach at parameter regions where analytical solutions are absent.
The composite three-component system is confined in an external harmonic trap and it is restricted in one-dimension. 
Due to the non-interacting nature of the impurities correlations among them are solely induced by their medium, whilst their dressing by the excitations of the bosonic host leads to the formation of Bose polarons.

As a first step, we categorize the emergent induced correlation regimes between the impurities, appearing in the system's ground state, for varying the individual impurity-medium couplings. 
In this sense, it is possible to associate that an induced correlated (anti-correlated) behavior as captured by the impurities integrated correlation function occurs for positive (negative) values of the product of the impurities-medium interactions.
The knowledge of the above correlated patterns serves as a guide for triggering the dynamics by using linear impurity-medium interaction ramps between distinct or within the same correlation regimes.

Specifically, in order to identify the effective mass and frequency of the Bose polaron we consider the case where one of the impurities interacts only in a mean-field manner with the medium while the other one can become correlated with its host. 
Here, three different one-body models were constructed.
The first is based on the assumption of adiabatically ramping up the interaction such that the system instantaneously follows its ground state configuration. 
Within the second model the host plays the role of an external time-dependent mean-field potential to the impurity. 
Finally, for the third reduced approach a time-dependent optimization routine is employed.
According to the latter, the time-dependent effective mass and frequency are fitting parameters ensuring that characteristic one-body observables predicted in the effective model match the ones obtained in the many-body method. 
This time-dependent optimized model leads to the most accurate description of the polaron characteristics.  

The estimation of each polaron effective mass and frequency is crucial for understanding the impurities induced correlations. This is accomplished by an effective two-body model which for each impurity combines the appropriate optimized effective one-body model and additionally includes a two-body interaction potential. 
The latter is modelled either through a time-dependent exponentially decaying interaction term or a contact interaction potential whose coupling is obtained either via an adiabatic interaction ramp or by following an optimization scheme.
We find that, similarly to the one-body case, the optimized effective two-body model provides the best agreement with the impurities time-evolved correlation function predicted within the many-body approach. 
Moreover, we showcase the presence of finite impurity-medium and impurity-impurity entanglement by calculating the corresponding von Neumann entropy and logarithmic negativity using a decomposition of the effective two-body impurities wave function.

Our results pave the way for future studies aiming to reduce the dynamics of a complex highly particle-imbalanced many-body system into effective one- or two-body models and ultimately engineer the polaron characteristics and induced correlations. 
This knowledge might be also proven useful for relevant extensions to higher dimensions. 
Possible future directions include, for instance, the treatment of mass-imbalanced three-component settings where either two heavy impurities are coupled to a lighter bosonic medium or the distinguishable impurities have different masses. 
Especially in the latter case, understanding the interplay of the effective mass with the induced interactions would be worth pursuing. 
Another interesting extension is to consider species selective 
trapping geometries such as a homogeneous or lattice trapped bath.

\section*{Acknowledgements} 

The authors would like to thank Ilias Englezos, Kapil Goswami and Ansgar Siemens for valuable discussions. 
S. I. M is supported from the Missouri Science and Technology, Department of Physics, Startup fund.

%%%%%%%%%%%%%%%%%%%%%%%%%%%%%%%%%%%%%%%%%%%%%%%%%%%%%%%%%%%%%%%%%%%%%%%%%%%%%%%%%%%%%%%%%%%%%%%
\appendix
%%%%%%%%%%%%%%%%%%%%%%%%%%%%%%%%%%%%%%%%%%%%%%%%%%%%%%%%%%%%%%%%%%%%%%%%%%%%%%%%%%%%%%%%%%%%%%%
\section{Impact of the ramp rate}
\label{app:ramp_time}

\begin{figure}
\centering
\includegraphics[width=1.0\linewidth]{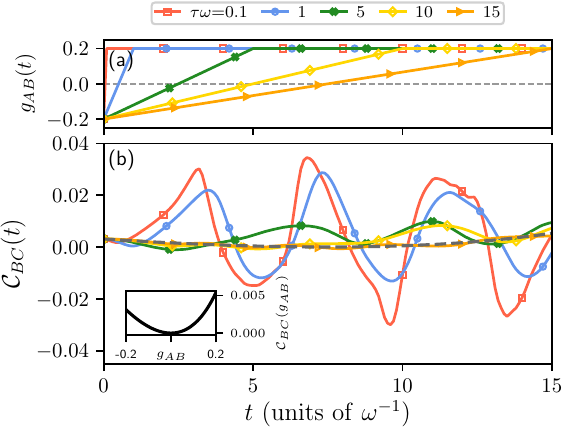}
\caption{(a) Dependence on the ramp rate of the linear impurity-medium interaction quench protocol from $g_{AB}^0=g_{AC}^0=-0.2$ to $g_{AB}^\tau=g_{AC}^\tau=0.2$. 
(b) Time-evolution of the impurities induced correlations described by their integrated correlation function $\mathcal{C}_{BC}(t)$ [Eq.~(\ref{eq:corr_int})]. 
Inset of panel (b) depicts the corresponding adiabatic evolution of $\mathcal{C}_{BC}(t)$ following the above protocol when  $\tau\rightarrow\infty$, which also appears as a gray dashed line in the main panel. 
An excellent agreement with the linear quench characterized by a rate $\tau \omega=15$ is observed.  
The system consists of two distinguishable non-interacting $B$ and $C$  impurities immersed in a bosonic medium with $N_A=10$ and $g_{AA}=0.2$.}
\label{fig:vary_ramp_t}
\end{figure}

In the following, we explicate the effect of the ramp rate, $\tau$, of the linear interaction quench [Eq.~(\ref{eq:interaction_ramp})] on the resulting induced correlation dynamics as captured by $\mathcal{C}_{BC}(t)$ [see Eq.~(\ref{eq:corr_int})]. For this investigation, we invoke a paradigmatic quench protocol namely ramping the impurity-medium couplings from $g_{AB}^0=g_{AC}^0=-0.2$ to $g_{AB}^\tau=g_{AC}^\tau=0.2$ for various $\tau$ as illustrated in Fig.~\ref{fig:vary_ramp_t}(a). Notice that a similar phenomenology occurs also for other post-quench interactions not shown for brevity. 

The emergent time-evolution of the impurities integrated correlation function, $\mathcal{C}_{BC}(t)$, is demonstrated in Fig.~\ref{fig:vary_ramp_t}(b).
As expected, we observe that relatively small ramp rates, i.e. $\tau \omega=0.1, 1$, lead to the most pronounced dynamical response of $\mathcal{C}_{BC}(t)$ manifested by its enhanced amplitude oscillatory behavior. 
However, induced correlations appear to be weaker when using longer ramp rates. 
A such, it is possible to infer that for an increasing quench rate, e.g. $\tau\omega=15$ in Fig.~\ref{fig:vary_ramp_t}(b),   
the correlation dynamics approaches its ideal adiabatic behavior which practically refers to $\tau\rightarrow\infty$. 
Indeed, during its adiabatic evolution the system remains always in its ground state and hence the respective  $\mathcal{C}_{BC}$ is equivalent to the corresponding static solution characterized by specific $g_{AB}$, $g_{AC}$ (or for the employed quench protocol only by $g_{AB}$). The $\mathcal{C}_{BC}$ of the ground state system configurations passing through the respective adiabatic evolution can be seen in the inset of Fig.~\ref{fig:vary_ramp_t}(b) and also in the main panel of Fig.~\ref{fig:vary_ramp_t}(b) as a dashed gray line. It becomes apparent that it nearly coincides with $\mathcal{C}_{BC}(t)$ for $\tau\omega=15$ further verifying the approach to the adiabatic limit. 
Therefore, the rates of $\tau\omega=1$ and $5$ which are employed in the main text induce a more prominent dynamical response that is closer to the sudden quench scenario.

%%%%%%%%%%%%%%%%%%%%%%%%%%%%%%%%%%%%%%%%%%%%%%%%%%%%%%%%%%%%%%%%%%%%%%%%%%%%%%%%%%%%%%%%%%%%%%%
\section{Determination of the ground state effective parameters}
\label{app:eff_model_gs}

\subsection{Effective mass and frequency of the one-body model}
\label{app:1b_model_gs}

The effective mass and frequency of the ground state Bose polaron~\cite{theel2024} are calculated as follows. 
The system of two distinguishable and non-interacting impurities coupled to a bosonic medium is numerically solved within a species mean-field ansatz (Sec.~\ref{sec:methodology}). 
The latter accounts only for the correlations of one impurity with the bath, while the other impurity acts as a mean-field potential.
Afterwards, we determine the effective mass and frequency of the impurity $\sigma=B$ ($C$) for a specific $(g_{AB}, g_{AC})$ combination. This is done by fitting the energy and position variance predicted by the effective one-body model 
\begin{align}
\hat{H}^{(1)}(m_\sigma^{\mathrm{eff}}, \omega_\sigma^{\mathrm{eff}}) = -\frac{\partial_x^2}{2 m_\sigma^{\rm{eff}}} + \frac{1}{2} m_\sigma^{\rm{eff}} \left(\omega_\sigma^{\rm{eff}}\right)^2 x^2,
\end{align}
to the ones evaluated by the suitable species mean-field ansatz. 
Thereby, the effective mass and frequency ($m_\sigma^{\mathrm{eff}}$, $\omega_\sigma^{\mathrm{eff}}$) are chosen such that the cost function $c_{\mathrm{gs}}^{1\mathrm{body}}=\Delta E_\sigma + \Delta x^2_{\sigma}$ becomes vanishingly small of the order of $10^{-9}$. 
Here, $\Delta E_\sigma = |E_\sigma^{\mathrm{sMF}\sigma'} - E_\sigma^{\mathrm{eff}}|^2$ is the difference between the energy of the effective one-body model, $E_\sigma^{\mathrm{eff}}$, and the energy of the impurity-bath system described by the appropriate species mean-field ansatz. Namely, $E_\sigma^{\mathrm{sMF}\sigma'}=\langle \Psi^{\mathrm{sMF}\sigma'}|\hat{h}_\sigma| \Psi^{\mathrm{sMF}\sigma'}\rangle$, where $\sigma'=C$ ($B$) and $\hat{h}_\sigma$ is the one-body Hamiltonian appearing in Eq.~(\ref{eq:MB_Hamiltonian}). Likewise, we define $\Delta x^2_\sigma = |\langle x_\sigma^2\rangle_{\mathrm{sMF}\sigma'} - \langle x^2_\sigma\rangle_{\mathrm{eff}}|^2$.

\subsection{Effective two-body interaction strength of the two-body model}
\label{app:2b_model_gs}

Having at hand the effective mass and frequency of the polaron it is possible to also find the impurities' effective interaction strength for each interaction configuration $(g_{AB}, g_{AC})$. 
Here, one should rely on the time-independent version of the effective two-body model of Eq.~(\ref{eq:2b_ham}). 
Then, the only free parameter is the effective contact interaction strength, $g_{BC}^{\rm{eff}}$, appearing in $\hat{V}_{BC}^{\rm{int}}(t)=g_{BC}^{\mathrm{eff}}\delta(x^B, x^C)$ [Eq.~(\ref{eq:2b_ham})].
The value of $g_{BC}^{\rm{eff}}$ is the one which minimizes the cost function $\Delta \mathcal{G}_{BC}=\int \int \rm{d}x^B\rm{d}x^C |\mathcal{G}_{BC}(x^B, x^C) - \mathcal{G}_{BC}^{\rm{eff,gs}}(x^B, x^C)|^2$. 
In this expression, $\mathcal{G}_{BC}^{\rm{eff,gs}}(x^B, x^C)$ ($\mathcal{G}_{BC}(x^B, x^C)$) is the impurity-impurity correlation function [Eq.~(\ref{eq:corr})] obtained with the effective (full many-body) approach. To calculate the static two-body solution in the effective model we expand the two-body wave function in terms of one-body states and diagonalize the respective two-body Hamiltonian (known as exact diagonalization method).

%%%%%%%%%%%%%%%%%%%%%%%%%%%%%%%%%%%%%%%%%%%%%%%%%%%%%%%%%%%%%%%%%%%%%%%%%%%%%%%%%%%%%%%%%%%%%%%
\section{Convergence behavior of the applied optimization routine}
\label{app:opt_convergence}
~
\begin{figure}
\centering
\includegraphics[width=1.0\linewidth]{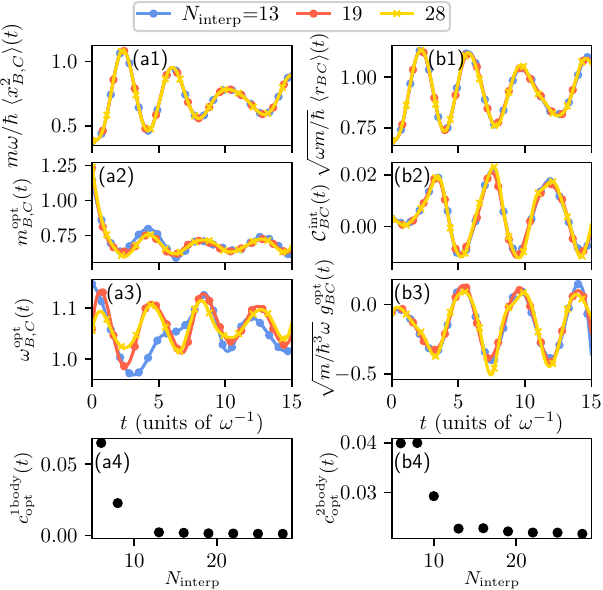}
\caption{Convergence study of different observables extracted from (a1)-(a4) the optimized one-body model as well as (b1)-(b4) the optimized two-body model. (a1)-(a3) Dynamical evolution of the variance, effective mass and frequency, respectively, for a varying number of interpolation points used in the optimization procedure (see Sec. \ref{sec:eff1b_opt}). (b1)-(b3) Time-evolution of the impurities' relative difference, integrated correlation function and effective induced interaction, respectively, obtained within the optimization routine described in Sec. \ref{sec:eff2b_opt}. (a4), (b4) Cost value of the optimized one- and two-body effective model, respectively, in dependence of the number of interpolation points. We consider two non-interacting impurities coupled to a bosonic medium with $N_A=10$ atoms interacting among each other with $g_{AA}=0.1$. The dynamics is induced by ramping the impurity-medium interaction strengths at the same time from $g_{AB}^0= g_{AC}^0=-0.2$ to $g_{AB}^\tau = g_{AC}^\tau = 0.2$ within a ramp time of $\tau\omega=1$.}
\label{fig:opt_convergence}
\end{figure}

The convergence behavior of the optimization routine used to determine the time-dependent effective mass and frequency of the polaron as well as the effective coupling of two distinguishable polarons (Sec.~\ref{sec:eff1b_opt}, \ref{sec:eff2b_opt}) depends on the number of interpolation points, $N_{\rm{interp}}$. 
This effect is demonstrated in Fig.~\ref{fig:opt_convergence} for linear interactions ramps from $g_{AB}^0= g_{AC}^0=-0.2$ to $g_{AB}^\tau = g_{AC}^\tau = 0.2$ within $\tau\omega=1$. 
In both optimization routines the aim is to approach the time-dependent path for which the observables of interest do not change, at least to a certain accuracy, by successively increasing $N_{\rm{interp}}$.
Simultaneously, the cost functions $c_{\mathrm{opt}}^{1\mathrm{body}}$ and $c_{\mathrm{opt}}^{2\mathrm{body}}$ (Sec.~\ref{sec:eff1b_opt}, \ref{sec:eff2b_opt}) should exhibit a decreasing trend for larger $N_{\rm{interp}}$.

Inspecting the convergence behavior of the optimized effective mass and frequency, shown in Fig.~\ref{fig:opt_convergence}(a2), (a3), we can infer numerical convergence of the individual observables with the optimization method for $N_{\rm{interp}}>19$. It is worth mentioning here  that the cost function $c_{\mathrm{opt}}^{1\mathrm{body}}$ saturates already for $N_{\rm{interp}}>12$ [Fig.~\ref{fig:opt_convergence}(a4)], and the same holds for the respective variance $\langle x_{B,C}^2 \rangle(t)$ in Fig. \ref{fig:opt_convergence}(a1). Thus, in such studies it is important to ensure convergence not only of the cost function but also of the effective parameters and the observables of interest.  
The same behavior can be also observed when inspecting the results obtained by optimizing the effective interaction strength, shown in Fig.~\ref{fig:opt_convergence}(b1)-(b4). However, here $c_{\mathrm{opt}}^{2\mathrm{body}}$ saturates at a finite value even upon considering a larger number of interpolation points as shown here. This implies that we have reached the limitation of the optimization procedure for higher-order observables.

\begin{figure*}
\centering
\includegraphics[width=0.95\linewidth]{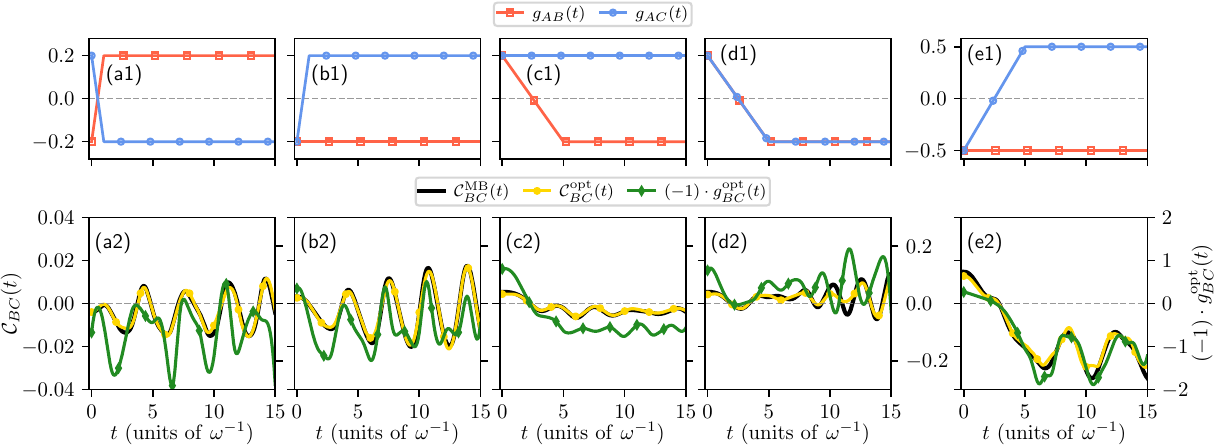}
\caption{
(a1)-(e1) Schematic representation of different linear impurity-medium interaction ramps characterized by rates (a1), (b1) $\tau\omega=1$, (c1)-(e1) $\tau\omega=5$. The initial $(g_{AB}^0, g_{AC}^0)=(\pm 0.2, \pm 0.2)$ and final $(g_{AB}^\tau, g_{AC}^\tau)=(\pm 0.2,\pm 0.2)$ interaction combinations are shown in the panels (a1)-(d1), while in panel (e1) they correspond to $(g_{AB}^0, g_{AC}^0)=(-0.5, -0.5)$ and $(g_{AB}^\tau, g_{AC}^\tau)=(-0.5,0.5)$.  
(a2)-(e2) Time-evolution of the impurities integrated two-body correlation function within different approaches following the interaction ramps illustrated in panels (a1)-(e1). 
For comparison, we provide the integrated correlation functions $\mathcal{C}_{BC}^{\mathrm{opt}}(t)$ and $\mathcal{C}_{BC}^{\mathrm{MB}}(t)$, obtained with the effective optimized two-body model and the full many-body approach (see legend). 
It becomes evident that the correlation function deduced from the optimization process of the two-body model leads to a very good agreement with the one from the many-body method.
In all cases, we consider two non-interacting distinguishable impurities $B$ and $C$ coupled to a bosonic medium of $N_A=10$ interacting atoms with $g_{AA}=0.2$.}
\label{fig:other_quenches_entanglement}
\end{figure*}

\section{Validity of the optimization method independently of the quench protocol}
\label{app:other_quench_protocols}

Let us demonstrate the applicability of the employed optimization routines (outlined in Sec.~\ref{sec:eff1b_opt}, \ref{sec:eff2b_opt}) irrespective of the used interaction ramp.
Exemplary time-dependent interaction ramps of $g_{AB}(t)$ and $g_{AC}(t)$ are illustrated in 
Fig.~\ref{fig:other_quenches_entanglement}(a1)-(e1).
The respective temporal-evolution of the integrated correlation function obtained from the full many-body method, $\mathcal{C}_{BC}^{\mathrm{MB}}(t)$, and the one calculated from the two-body optimization routine with the aid of the reduced two-body model described in Sec.~\ref{sec:eff2b_opt}, $\mathcal{C}_{BC}^{\mathrm{opt}}(t)$, is presented in Fig.~\ref{fig:other_quenches_entanglement}(a2)-(e2). Of course, preceding this two-body optimization routine, we have estimated the underlying time-dependent effective mass and frequencies as discussed in Sec.~\ref{sec:eff1b_opt} (not shown).
To ease the visualization of the above observables we invert the sign of $g_{BC}^{\rm{opt}}(t)$ such that a positive (negative) value of $\mathcal{C}_{BC}^{\mathrm{opt}}$, associated with correlated (anti-correlated) impurities is accompanied by negative (positive) values of $-g_{BC}^{\rm{opt}}(t)$. 
The quality of the optimization routines is judged by comparing  $\mathcal{C}_{BC}^{\mathrm{opt}}(t)$ and $\mathcal{C}_{BC}^{\mathrm{MB}}(t)$.
Overall, we find a very good agreement which validates the outcomes of the effective one- and two-body models. 
We finally remark that in the course of the dynamics quench-induced excitation patterns may be imprinted in the correlation function whose integration could oversimplify such a complex pattern. 
Still, in the considered cases we find that the integrated correlation function provides a reliable estimate about the impurities induced correlation behavior.

%%%%%%%%%%%%%%%%%%%%%%%%%%%%%%%%%%%%%%%%%%%%%%%%%%%%%%%%%%%%%%%%%%%%%%%%%%%%%%%%%%%%%%%%%%%%%
\section{Decomposing the two-body dynamics}
\label{app:decomposing_eff2b}

\begin{figure}
\centering
\includegraphics[width=1.0\linewidth]{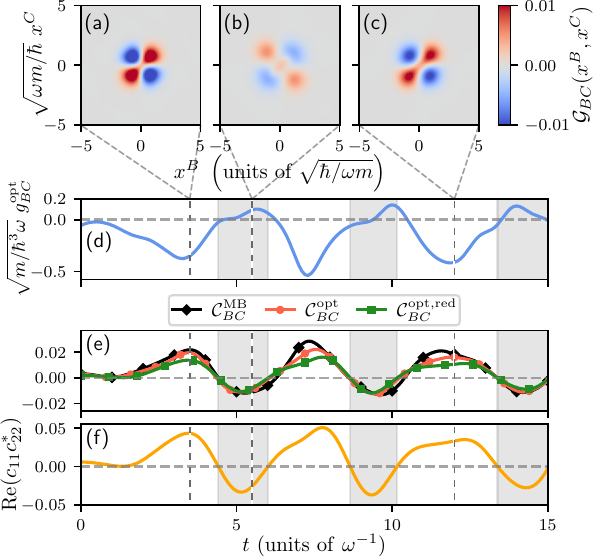}
\caption{(a)-(c) Snapshots of the impurity-impurity correlation function, $\mathcal{G}_{BC}(x^B, x^C)$, at times $t\omega=3.5,5.5,12$. 
(d) Time-evolution of the effective impurity-impurity induced interaction, $g_{BC}^{\rm{opt}}$, retrieved by fitting a two-body model to the many-body dynamics (Sec.~\ref{sec:eff2b_opt}).
(e) Integrated correlation function obtained within the many-body dynamics $\mathcal{C}^{\rm{MB}}_{BC}$ and the effective two-body model $\mathcal{C}^{\rm{opt}}_{BC}$ compared to the reduced optimized integrated correlation function $\mathcal{C}^{\rm{opt, red}}_{BC}$.
(f) Real part of the product of expansion coefficients $c_{11}c_{22}^*$ signifying the decisive role of the correlation expansion coefficients in the interpretation of the two-body correlation behavior. 
In all cases, the two non-interacting impurities $B$ and $C$ are coupled to the bosonic medium with $N_A=10$ and $g_{AA}=0.2$. The dynamics is induced by a linear ramp of the impurity-medium coupling from $g_{AB}^0=g_{AC}^0=-0.2$ to $g_{AB}^\tau=g_{AC}^\tau=0.2$ with rate $\tau \omega = 1$.}
\label{fig:decomposition}
\end{figure}

To gain a deeper understanding on the participating microscopic processes responsible for the observed induced impurity-impurity correlation dynamics we shall next carefully inspect the ingredient of the correlation function.
As a characteristic case example we focus on interaction ramps from $g_{AB}^0=g_{AC}^0=-0.2$ to $g_{AB}^\tau=g_{AC}^\tau=0.2$ with rate $\tau \omega = 1$.
As it will be showcased below, reducing the ramp rate to $\tau \omega = 1$ triggers a more complex response where induced correlations switch from a correlated to an anti-correlated behavior and vice versa even for $\tau\leq t$. 
As such, this allows to provide a more general argumentation for the behavior of correlations.
Snapshots of $\mathcal{G}_{BC}(x^B, x^C; t)$ within the full many-body method at short evolution times are depicted in Fig.~\ref{fig:decomposition}(a)-(c). 
Also, the respective time-evolution of the integrated correlation, $\mathcal{C}^{\rm{MB}}_{BC}(t)$, is illustrated in Fig.~\ref{fig:decomposition}(e).  
The oscillatory trend of $\mathcal{C}^{\rm{MB}}_{BC}(t)$ taking positive and negative values indicates the periodic appearance of correlated and anti-correlated patterns which are also evident in the spatially resolved correlation provided in Fig.~\ref{fig:decomposition}(a)-(c). 
We remark that such an oscillatory correlation behavior can not be observed for larger ramp rates, see Fig.~\ref{fig:2b_model_res}(b2) and Appendix~\ref{app:ramp_time} further justifying our choice for the rate.

To track the induced impurities interactions we deploy their effective two-body optimized description [Sec.~\ref{sec:eff2b_opt}] where the underlying interaction potential contains a time-dependent interaction strength as shown in Fig.~\ref{fig:decomposition}(d). 
It can be seen that the resulting integrated correlation function, $\mathcal{C}^{\rm{opt}}_{BC}(t)$ [Fig.~\ref{fig:decomposition}(e)] agrees qualitatively well with $\mathcal{C}^{\rm{MB}}_{BC}(t)$. 
Hence, this effective two-body model captures the aforementioned alternating induced correlation behavior traced back to the alternating sign of the induced interaction strength $g_{BC}^{\rm{opt}}(t)$, see Fig.~\ref{fig:decomposition}(d).

The effective two-body wave function can be expressed in terms of the time-dependent single-particle basis of harmonic oscillator eigenstates $|\phi_i^{B}(t)\rangle$ and $|\phi_j^{C}(t)\rangle$. The latter are determined through the optimization scheme yielding the time-dependent polaron effective masses and frequencies, i.e., $m_{\sigma}^{\rm{opt}}(t)$ and $\omega_{\sigma}^{\rm{opt}}(t)$. As such  
\begin{align}
|\Psi_{BC}^{\mathrm{opt}}(t) \rangle = \sum_{ij} c_{ij}(t) |\phi_i^{B}(t)\rangle \otimes |\phi_j^{C}(t)\rangle.
\end{align}
Utilizing this wave function expansion the impurities' two-body optimized correlation function can be approximated as
\begin{align}
\mathcal{G}_{BC}^{\mathrm{opt}}(t) &\approx \sum_{ijkl} c_{ij}(t)c_{kl}^*(t) \mathcal{G}^{\mathrm{opt}}_{ijkl}
\end{align}
with
\begin{align}
\mathcal{G}^{\mathrm{opt}}_{ijkl} =& \rho^{(2),\mathrm{opt}}_{ijkl}(t)- \int \rho^{(2),\mathrm{opt}}_{ijkl}(t) \mathrm{d}x^C \int \rho^{(2),\mathrm{opt}}_{ijkl}(t) \mathrm{d}x^B.
\label{2bopt_correl}
\end{align}
Here, the matrix elements of the two-body reduced density matrix are given by $\rho^{(2),\mathrm{opt}}_{ijkl}(t)= \phi_i^{B}(t) \phi_j^{C}(t) \phi_k^{B,\dagger}(t) \phi_l^{C,\dagger}(t)$
and, for simplicity, we have dropped the spatial coordinates $x^B$ and $x^C$. 
Also, the last term in Eq.~(\ref{2bopt_correl}) is extracted in order to eliminate the unconditional probability [see also Eq.~(\ref{eq:corr})]. 
Notice that by doing this individually in each expansion term $\rho^{(2),\mathrm{opt}}_{ijkl}(t)$ of the actual two-body density is already an approximation for $\mathcal{G}_{BC}^{\mathrm{opt}}(t)$  since we obtain the unconditional probability individually for each expansion term instead of retrieving it from the complete two-body density.  
It turns out that within this simplification
\begin{align}
\mathcal{G}_{ijkl}^{\mathrm{opt}}(t) =
\begin{cases}
0, & i=k~\textrm{and}~j=l, \\
\rho^{(2),\mathrm{opt}}_{ijkl}(t), &\text{else,}
\end{cases}  
\end{align}
which is a direct consequence of the orthonormality relation of the one-body basis functions.
Further integrating $\mathcal{G}_{ijkl}^{\mathrm{opt}}(t)$ according to Eq.~(\ref{eq:corr_int}) yields  $\mathcal{C}_{ijkl}^{\mathrm{opt}}(t)$. 
Here, we numerically identify certain $ijkl$ combinations for which the elements of $\mathcal{C}_{ijkl}^{\mathrm{opt}}(t)$ are suppressed, e.g., for $ijkl=1113$ where the positive and negative portions of the orbitals building up $\rho^{(2),\mathrm{opt}}_{ijkl}$ cancel each other when integrating. 
Taking all these findings into account, we find that a good approximation for $\mathcal{C}_{BC}^{\mathrm{opt}}(t)$ is represented by
\begin{align}
\mathcal{C}_{BC}^{\mathrm{opt, red}}(t) = 
 2\rm{Re}[c_{11}(t)c_{22}^*(t)]\mathcal{C}_{1122}.
\end{align}
To validate these approximations, the reduced integrated two-body correlation function $\mathcal{C}_{BC}^{\mathrm{opt, red}}(t)$ is shown together with the one obtained from the optimization process in Fig.~\ref{fig:decomposition}(e). A direct comparison reveals a good qualitative agreement between the two, thus verifying our arguments. 

Importantly, within this approximative picture it is possible to trace the origin of the alternating sign of the impurities induced interaction and correlations. This is explained by the time-evolution of the $c_{11}(t)c_{22}^*(t)$ coefficients which is presented in  Fig.~\ref{fig:decomposition}(f). 
In particular, notice the gray shaded areas in Fig.~\ref{fig:decomposition}(d)-(f) marking the time-intervals where $\rm{Re}[c_{11}(t)c_{22}^*(t)]<0$ which correspond to an anti-correlated impurities behavior and repulsive induced interactions. 

\bibliography{project_dtm.bib}

\end{document}